# Properties of the Binary Neutron Star Merger GW170817

B. P. Abbott *et al.*[*]
(LIGO Scientific Collaboration and Virgo Collaboration)

On August 17, 2017, the Advanced LIGO and Advanced Virgo gravitational-wave detectors observed a low-mass compact binary inspiral. The initial sky localization of the source of the gravitational-wave signal, GW170817, allowed electromagnetic observatories to identify NGC 4993 as the host galaxy. In this work, we improve initial estimates of the binary's properties, including component masses, spins, and tidal parameters, using the known source location, improved modeling, and recalibrated Virgo data. We extend the range of gravitational-wave frequencies considered down to 23 Hz, compared to 30 Hz in the initial analysis. We also compare results inferred using several signal models, which are more accurate and incorporate additional physical effects as compared to the initial analysis. We improve the localization of the gravitational-wave source to a 90% credible region of 16 $\mathrm{deg}^2$. We find tighter constraints on the masses, spins, and tidal parameters, and continue to find no evidence for nonzero component spins. The component masses are inferred to lie between 1.00 and 1.89 $\mathrm{M}_\odot$ when allowing for large component spins, and to lie between 1.16 and 1.60 $\mathrm{M}_\odot$ (with a total mass $2.73^{+0.04}_{-0.01}$ $\mathrm{M}_\odot$) when the spins are restricted to be within the range observed in Galactic binary neutron stars. Using a precessing model and allowing for large component spins, we constrain the dimensionless spins of the components to be less than 0.50 for the primary and 0.61 for the secondary. Under minimal assumptions about the nature of the compact objects, our constraints for the tidal deformability parameter $\tilde{\Lambda}$ are (0,630) when we allow for large component spins, and $300^{+420}_{-230}$ (using a 90% highest posterior density interval) when restricting the magnitude of the component spins, ruling out several equation-of-state models at the 90% credible level. Finally, with LIGO and GEO600 data, we use a Bayesian analysis to place upper limits on the amplitude and spectral energy density of a possible postmerger signal.

## I. INTRODUCTION

On August 17, 2017, the advanced gravitational-wave (GW) detector network, consisting of the two Advanced LIGO detectors [1] and Advanced Virgo [2], observed the compact binary inspiral event GW170817 [3] with a total mass less than any previously observed binary coalescence and a matched-filter signal-to-noise ratio (SNR) of 32.4, louder than any signal to date. Follow-up Bayesian parameter inference allowed GW170817 to be localized to a relatively small sky area of 28 $\mathrm{deg}^2$ and revealed component masses consistent with those of binary neutron star (BNS) systems. In addition, 1.7 s after the binary's coalescence time, the Fermi and INTEGRAL gamma-ray telescopes observed the gamma-ray burst GRB 170817A with an inferred sky location consistent with that measured for GW170817 [4], providing initial evidence that the binary system contained neutron star (NS) matter.

Astronomers followed up on the prompt alerts produced by this signal, and within 11 hours the transient SSS17a/AT 2017gfo was discovered [5,6] and independently observed by multiple instruments [7–11], localizing the source of GW170817 to the galaxy NGC 4993. The identification of the host galaxy drove an extensive follow-up campaign [12], and analysis of the fast-evolving optical, ultraviolet, and infrared emission was consistent with that predicted for a kilonova [13–17] powered by the radioactive decay of $r$-process nuclei synthesized in the ejecta (see Refs. [18–28] for early analyses). The electromagnetic (EM) signature, observed throughout the entire spectrum, provides further evidence that GW170817 was produced by the merger of a BNS system (see, e.g., Refs. [29–31]).

According to general relativity, the gravitational waves emitted by inspiraling compact objects in a quasicircular orbit are characterized by a chirplike time evolution in their frequency that depends primarily on a combination of the component masses called the chirp mass [32] and secondarily on the mass ratio and spins of the components. In contrast to binary black hole (BBH) systems, the internal structure of the NS also impacts the waveform and needs to be included for a proper description of the binary evolution. The internal structure can be probed primarily through attractive tidal interactions that lead to an accelerated inspiral. These tidal interactions are small at lower GW frequencies but increase rapidly in strength during the final tens of GW cycles before merger. Although tidal effects are small relative to other effects, their distinct behavior makes them potentially measurable from the GW signal [33–37], providing additional evidence for a BNS system and insight into the internal structure of NSs.

[*]Full author list given at the end of the article.

In this work, we present improved constraints on the binary parameters first presented in Ref. [3]. These improvements are enabled by (i) recalibrated Virgo data [38], (ii) a broader frequency band of 23–2048 Hz as compared to the original 30–2048 Hz band used in Ref. [3], (iii) a wider range of more sophisticated waveform models (see Table I), and (iv) knowledge of the source location from EM observations. By extending the bandwidth from 30–2048 Hz to 23–2048 Hz, we gain access to an additional about 1500 waveform cycles compared to the about 2700 cycles in the previous analysis. Overall, our results for the parameters of GW170817 are consistent with, but more precise than, those in the initial analysis [3]. The main improvements are (i) improved 90% sky localization from 28 $\mathrm{deg}^2$ to 16 $\mathrm{deg}^2$ without use of EM observations, (ii) improved constraint on inclination angle enabled by independent measurements of the luminosity distance to NGC 4993, (iii) limits on precession from a new waveform model that includes both precession and tidal effects, and (iv) evidence for a nonzero tidal deformability parameter that is seen in all waveform models. Finally, we analyze the potential postmerger signal with an unmodeled Bayesian inference method [39] using data from the Advanced LIGO detectors and the GEO600 detector [40]. This method allows us to place improved upper bounds on the amount of postmerger GW emission from GW170817 [41].

As in the initial analysis of GW170817 [3], we infer the binary parameters from the inspiral signal while making minimal assumptions about the nature of the compact objects, in particular, allowing the tidal deformability of each object to vary independently. We allow for a large range of tidal deformabilities in our analysis, including zero, which means that our analysis includes the possibility of phase transitions within the stars and allows for exotic compact objects or even for black holes as binary components. In a companion paper [42], we present a complementary analysis assuming that both compact objects are NSs obeying a common equation of state (EOS). This analysis results in stronger constraints on the tidal deformabilities of the NSs than we can make under our minimal assumptions, and it allows us to constrain the radii of the NSs and make for novel inferences about the equation of state of cold matter at supranuclear densities.

This paper is organized as follows. Section II details the updated analysis, including improvements to the instrument calibration, improved waveform models, and additional constraints on the source location. Section III reports the improved constraints on the binary's sky location, inclination angle, masses, spins, and tidal parameters. Section IV provides upper limits on possible GW emission after the binary merger. Finally, Sec. V summarizes the results and highlights remaining work such as inference of the NS radius and EOS. Additional results from a range of waveform models are reported in Appendix A, and an injection and recovery study investigating the systematic errors in our waveform models is given in Appendix B.

## II. METHODS

### A. Bayesian method

All available information about the source parameters $\vec{\vartheta}$ of GW170817 can be expressed as a posterior probability density function (PDF) $p(\vec{\vartheta}|d(t))$ given the data $d(t)$ from the GW detectors. Through application of Bayes' theorem, this posterior PDF is proportional to the product of the likelihood $\mathcal{L}(d(t)|\vec{\vartheta})$ of observing data $d(t)$ given the waveform model described by $\vec{\vartheta}$ and the prior PDF $p(\vec{\vartheta})$ of observing those parameters. Marginalized posteriors are computed through stochastic sampling using an implementation of Markov-chain Monte Carlo algorithm [43,44] available in the LALINFERENCE package [45] as part of the LSC Algorithm Library (LAL) [46]. By marginalizing over all but one or two parameters, it is then possible to generate credible intervals or credible regions for those parameters.

### B. Data

For each detector, we assume that the noise is additive, i.e., a data stream $d(t) = h^M(t;\vec{\vartheta}) + n(t)$ where $h^M(t;\vec{\vartheta})$ is the measured gravitational-wave strain and $n(t)$ is Gaussian and stationary noise characterized by the one-sided power spectral densities (PSDs) shown in Fig. 1. The PSD is defined as $S_n \equiv (2/T)\langle|\tilde{n}(f)|^2\rangle$, where $\tilde{n}(f)$ is the Fourier transform of $n(t)$ and the angle brackets indicate a time average over the duration of the analysis $T$, in this case the 128 s containing the $d(t)$ used for all results presented in Sec. III. This PSD is modeled as a cubic spline for the broadband structure and a sum of Lorentzians for the line features, using the BAYESWAVE package [39,47], which produces a posterior PDF of PSDs. Here, we approximate the full structure and variation of these posteriors as a point estimate by using a median PSD, defined separately at each frequency.

The analyses presented here use the same data and calibration model for the LIGO detectors as Ref. [3], including subtraction of the instrumental glitch present in LIGO-Livingston (cf. Fig. 2 of Ref. [3]) and of other independently measured noise sources as described in

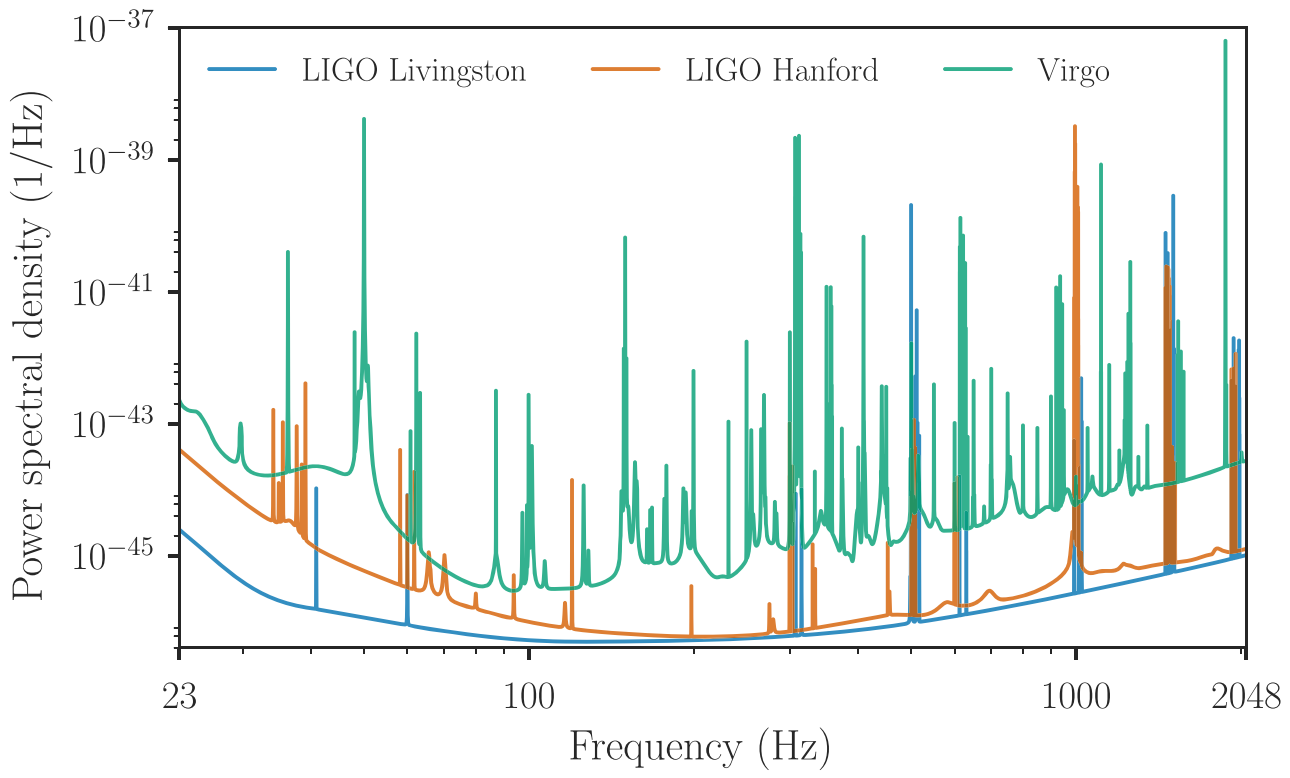

FIG. 1. PSDs of the Advanced LIGO–Advanced Virgo network. Shown, for each detector, is the median PSD computed from a posterior distribution of PSDs as estimated by BayesWave [39,47] using 128 s of data containing the signal GW170817.

Refs. [48–51]. The method used for subtracting the instrumental glitch leads to unbiased parameter recovery when applied to simulated signals injected on top of similar glitches in detector data [52]. The data from Virgo has been recalibrated since the publication of Ref. [3], including the subtraction of known noise sources during postprocessing of the data, following the procedure of Ref. [38] (the same as described in Sec. II of Ref. [53]). While the assumption of stationary, Gaussian noise in the detectors is not expected to hold over long timescales, our subtraction of the glitch, known noise sources, and recalibration of the Virgo data helps bring the data closer to this assumption. Applying the Anderson-Darling test to the data whitened by the on-source PSDs generated by BayesWave, we do not reject the null hypothesis that the whitened data are consistent with zero-mean, unit-variance, Gaussian noise $\mathcal{N}(0,1)$. The test returns $p$-values $> 0.1$ for the LIGO detectors' data. Meanwhile, the test is marginal when applied to the Virgo data, with $p \sim 0.01$. However, the information content of the data is dominated by the LIGO detectors, as they contained the large majority of the recovered signal power. The results of the Anderson-Darling tests support the use of the likelihood function as described in Ref. [45] for the signal characterization analyses reported in this paper.

The measured strain $h^M(t;\vec{\vartheta})$ may differ from the true GW strain $h(t;\vec{\vartheta})$ due to measured uncertainties in the detector calibration [54,55]. We relate the measured strain to the true GW strain with the expression [56,57]

$$\tilde{h}^M(f;\vec{\vartheta}) = \tilde{h}(f;\vec{\vartheta})[1+\delta A(f;\vec{\theta}^{\mathrm{cal}})]\exp[i\delta\phi(f;\vec{\theta}^{\mathrm{cal}})], \quad (1)$$

where $\tilde{h}^M(f;\vec{\vartheta})$ and $\tilde{h}(f;\vec{\vartheta})$ are the Fourier transforms of $h^M(t;\vec{\vartheta})$ and $h(t;\vec{\vartheta})$, respectively. The terms $\delta A(f;\vec{\theta}^{\mathrm{cal}})$ and $\delta\phi(f;\vec{\theta}^{\mathrm{cal}})$ are the frequency-dependent amplitude correction factor and phase correction factor, respectively, and are each modeled as cubic splines. For each detector, the parameters are the values of $\delta A$ and $\delta\phi$ at each of ten spline nodes spaced uniformly in $\log f$ [58] between 23 and 2048 Hz.

For the LIGO detectors, the calibration parameters $\vec{\theta}^{\mathrm{cal}}$ are informed by direct measurements of the calibration uncertainties [54] and are modeled in the same way as in Ref. [3] with $1\sigma$ uncertainties of less than 7% in amplitude and less than 3 degrees in phase for LIGO Hanford and less than 5% in amplitude and less than 2 degrees in phase for LIGO Livingston, all allowing for a nonzero mean offset. The corresponding calibration parameters for Virgo follow Ref. [38], with a $1\sigma$ amplitude uncertainty of 8% and a $1\sigma$ phase uncertainty of 3 degrees. This is supplemented with an additional uncertainty in the time stamping of the data of 20 $\mu$s (to be compared to the LIGO timing uncertainty of less than 1 $\mu$s [59] already included in the phase correction factor). At each of the spline nodes, a Gaussian prior is used with these $1\sigma$ uncertainties and their corresponding means. By sampling these calibration parameters in addition to the waveform parameters, the calibration uncertainty is marginalized over. This marginalization broadens the localization posterior (Sec. III A) but does not significantly affect the recovered masses, spins, or tidal deformability parameters.

TABLE I. Waveform models employed to measure the source properties of GW170817. The models differ according to how they treat the inspiral in the absence of tidal corrections [i.e., the BBH-baseline—in particular, the point particle (PP), spin-orbit (SO), and spin-spin (SS) terms], the manner in which tidal corrections are applied, whether the spin-induced quadrupoles of the neutron stars [67–70,79] are incorporated, and whether the model allows for precession or only treats aligned spins. Our standard model, PhenomPNRT, incorporates effective-one-body (EOB)- and numerical relativity (NR)-tuned tidal effects, the spin-induced quadrupole moment, and precession.

| Model name | Name in LALSuite | BBH baseline | Tidal effects | Spin-induced quadrupole effects | Precession |
|---|---|---|---|---|---|
| TaylorF2 | TaylorF2 | 3.5PN (PP [60–65], SO [66] SS [67–70]) | 6PN [71] | None | ✗ |
| SEOBNRT | SEOBNRv4_ROM_NRTidal | SEOBNRv4_ROM [72,73] | NRTidal [74,75] | None | ✗ |
| PhenomDNRT | IMRPhenomD_NRTidal | IMRPhenomD [76,77] | NRTidal [74,75] | None | ✗ |
| PhenomPNRT | IMRPhenomPv2_NRTidal | IMRPhenomPv2 [78] | NRTidal [74,75] | 3PN [67–70,79] | ✓ |

### C. Waveform models for binary neutron stars

In this paper, we use four different frequency-domain waveform models which are fast enough to be used as templates in LALINFERENCE. These waveforms incorporate point-particle, spin, and tidal effects in different ways. We briefly describe them below. Each waveform's key features are stated in detail in Table I, and further tests of the performance of the waveform models can be found in Ref. [75]. In addition to these frequency domain models, we employ two state-of-the-art time-domain tidal EOB models that also include spin and tidal effects [80,81]. These tidal EOB models have shown good agreement in comparison with NR simulations [80–83] in the late inspiral and improve on the post-Newtonian (PN) dynamics in the early inspiral. However, these implementations are too slow for use in LALINFERENCE. We describe these models in Sec. III D when we discuss an alternative parameter-estimation code [84,85].

The TaylorF2 model used in previous work is a purely analytic PN model. It includes point-particle and aligned-spin terms to 3.5PN order as well as leading-order (5PN) and next-to-leading-order (6PN) tidal effects [34,60–71,86–88]. The other three waveform models begin with point-particle models and add a fit to the phase evolution from tidal effects, labeled NRTidal [74,75], that fit the high-frequency region to both an analytic EOB model [80] and NR simulations [74,89]. The SEOBNRT model is based on the aligned-spin point-particle EOB model presented in Ref. [72] using methods presented in Ref. [73] to allow fast evaluation in the frequency domain. PhenomDNRT is based on an aligned-spin point-particle model [76,77] calibrated to untuned EOB waveforms [90] and NR hybrids [76,77]. Finally, PhenomPNRT is based on the point-particle model presented in Ref. [78], which includes an effective description of precession effects. In addition to tidal effects, PhenomPNRT also includes the spin-induced quadrupole moment that enters in the phasing at 2PN order [91]. For aligned-spin systems, PhenomPNRT differs from PhenomDNRT only in the inclusion of the spin-induced quadrupole moment. We include the EOS dependence of each NS's spin-induced quadrupole moment by relating it to the tidal parameter of each NS using the quasi-universal relations of Ref. [92]. Although this 2PN effect can have a large phase contribution, even for small spins [37], it enters at similar PN order as many other terms. We therefore expect it to be degenerate with the mass ratio and spins.

These four waveform models have been compared to waveforms constructed by hybridizing BNS EOB inspiral waveforms [80,83] with NR waveforms [74,80,82,89] of the late inspiral and merger. Since only the PhenomPNRT model includes the spin-induced quadrupole moment, it was found that it has smaller mismatches than PhenomDNRT and SEOBNRT [75]. In addition, because PhenomPNRT is the only model that includes precession effects, we use it as our reference model throughout this paper.

In Fig. 2, we show differences in the amplitude and phase evolution between the four models for an equal-mass, nonspinning BNS system. The top panel shows the fractional difference in the amplitude $\Delta A/A$ between each model and PhenomPNRT, while the bottom panel shows the absolute phase difference $|\Delta\Phi|$ between each model and PhenomPNRT. Because none of the models have amplitude corrections from tidal effects, the amplitude differences between the models are entirely due to the underlying point-particle models. For the nonprecessing system shown here, PhenomPNRT and PhenomDNRT agree by construction, and the difference with SEOBNRT is also small. On the other hand, the purely analytic TaylorF2 model that has not been tuned to NR simulations deviates by up to 30% from the other models. For the phase evolution of nonspinning systems, PhenomDNRT, PhenomPNRT, and SEOBNRT have the same tidal prescription, so the small $\lesssim 2$ rad phase differences are due to the underlying point-particle models. For nonspinning systems PhenomDNRT and PhenomPNRT are the same, but for spinning systems, the spin-induced quadrupole moment included in PhenomPNRT but not in PhenomDNRT will cause an additional phase difference. For TaylorF2 the difference with respect to PhenomPNRT is due to both the underlying point-particle model and the tidal prescription, and it is about 5 rad for nonspinning systems.

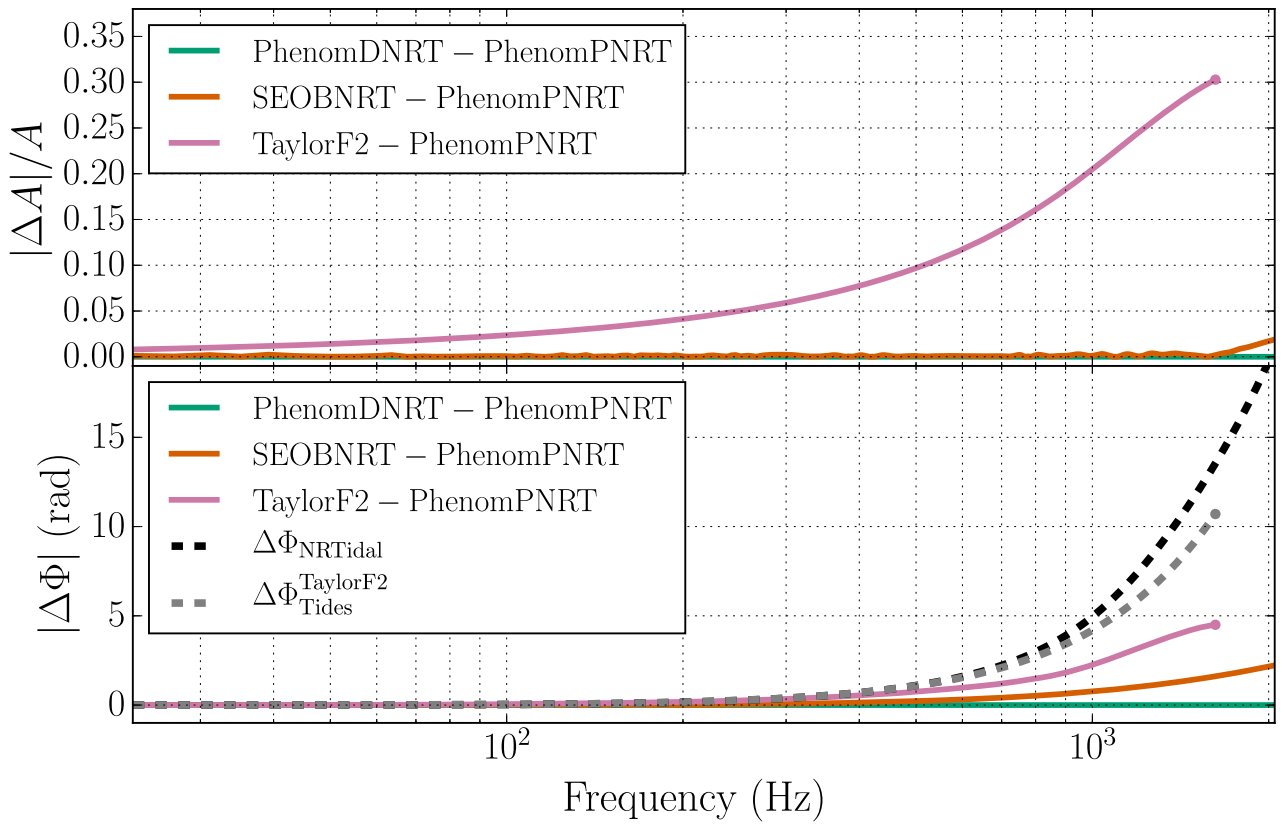

FIG. 2. Relative amplitude and phase of the employed waveform models starting at 23 Hz (see Table I) with respect to PhenomPNRT after alignment within the frequency interval [30, 30.25] Hz. Note that, in particular, the alignment between SEOBNRT and PhenomPNRT is sensitive to the chosen interval due to the difference in the underlying BBH-baseline models at early frequencies. We show, as an example, an equal-mass, nonspinning system with a total mass of 2.706 $M_\odot$ and a tidal deformability of $\tilde{\Lambda} = 400$. In the bottom panel, we also show the tidal contribution to the phasing for the TaylorF2 and the NRTidal models. This contribution can be interpreted as the phase difference between the tidal waveform models and the corresponding BBH models. The TaylorF2 waveform terminates at the frequency of the innermost stable circular orbit, which is marked by a small dot.

For reference, we also show in Fig. 2 the tidal contribution to the phase for the NRTidal models ($\Delta\Phi_{\rm NRTidal}$) and the TaylorF2 model ($\Delta\Phi^{\rm TaylorF2}_{\rm Tides}$). For the system here with tidal deformability $\tilde{\Lambda} = 400$ [Eq. (5)], the tidal contribution is larger than the differences due to the underlying point-particle models.

### D. Source parameters and choice of priors

The signal model for the quasicircular inspiral of compact binaries is described by intrinsic parameters that describe the components of the binary and extrinsic parameters that determine the location and orientation of the binary with respect to the observer. The intrinsic parameters include the component masses $m_1$ and $m_2$, where we choose the convention $m_1 \geq m_2$. The best measured parameter for systems displaying a long inspiral is the chirp mass [32,61,93,94],

$$\mathcal{M} = \frac{(m_1 m_2)^{3/5}}{(m_1 + m_2)^{1/5}}. \tag{2}$$

Meanwhile, ground-based GW detectors actually measure redshifted (detector-frame) masses, and these are the quantities we state our prior assumptions on. Detector-frame masses are related to the astrophysically relevant source-frame masses by $m^{\rm det} = (1+z)m$, where $z$ is the redshift of the binary [93,95]. Dimensionless quantities such as the ratio of the two masses, $q = m_2/m_1 \leq 1$, are thus the same in the detector frame and the source frame. When exploring the parameter space $\vec{\vartheta}$, we assume a prior PDF $p(\vec{\vartheta})$ uniform in the detector-frame masses, with the constraint that $0.5\ \mathrm{M}_\odot \leq m_1^{\rm det},\ m_2^{\rm det} \leq 7.7\ \mathrm{M}_\odot$, where $m_1^{\rm det} \geq m_2^{\rm det}$, and with an additional constraint on the chirp mass, $1.184\ \mathrm{M}_\odot \leq \mathcal{M}^{\rm det} \leq 2.168\ \mathrm{M}_\odot$. These limits were chosen to mimic the settings in Ref. [3] to allow for easier comparisons and were originally selected for technical reasons. The posterior does not have support near those limits. Despite correlations with the prior on the distance to the source, the source masses also have an effectively uniform prior in the region of parameter space relevant to this analysis.

When converting from detector-frame to source-frame quantities, we use the MUSE/VLT measurement of the heliocentric redshift of NGC 4993, $z_{\rm helio} = 0.0098$, reported in Refs. [96,97]. We convert this into a geocentric redshift using the known time of the event, yielding $z = 0.0099$.

The spin angular momenta of the two binary components $\mathbf{S}_i$ represent six additional intrinsic parameters and are usually represented in their dimensionless forms $\boldsymbol{\chi}_i = c\mathbf{S}_i/(Gm_i^2)$. For these parameters, we have, following Ref. [3], implemented two separate priors for the magnitudes of the dimensionless spins, $|\boldsymbol{\chi}| = \chi$, of the two objects. In both cases, we assume isotropic and uncorrelated orientations for the spins, and we use a uniform prior for the spin magnitudes, up to a maximum magnitude. In the first case, we enforce $\chi \leq 0.89$ to be consistent with the value used in Ref. [3]. This allows us to explore the possibility of exotic binary systems. The exact value of this upper limit does not significantly affect the results. Meanwhile, observations of pulsars indicate that, while the fastest-spinning neutron star has an observed $\chi \lesssim 0.4$ [98], the fastest-spinning BNSs capable of merging within a Hubble time, PSR J0737–3039A [99] and PSR J1946+2052 [100], will have at most dimensionless spins of $\chi \sim 0.04$ or $\chi \sim 0.05$ when they merge. Consistent with this population of BNS systems, in the second case we restrict $\chi \leq 0.05$.

For the waveforms in Table I that do not support spin precession, the components of the spins aligned with the orbital angular momentum $\chi_{1z}$ and $\chi_{2z}$ still follow the same prior distributions, which are marginalized over the unsupported spin components. We use the labels "high-spin" and "low-spin" to refer to analyses that use the prior $\chi \leq 0.89$ and $\chi \leq 0.05$, respectively.

The dimensionless parameters $\Lambda_i$ governing the tidal deformability of each component, discussed in greater detail in Sec. III D, are given a prior distribution uniform within $0 \leq \Lambda_i \leq 5000$, where no correlation between $\Lambda_1$, $\Lambda_2$, and the mass parameters is assumed. If we assume the two components are NSs that obey the same EOS, then $\Lambda_1$ and $\Lambda_2$ must have similar values when $m_1$ and $m_2$ have similar values [101–103]. This additional constraint is discussed in a companion paper that focuses on the NS EOS [42].

The remaining signal parameters in $\vec{\vartheta}$ are extrinsic parameters that give the localization and orientation of the binary. When we infer the location of the binary from GW information alone (in Sec. III A), we use an isotropic prior PDF for the location of the source on the sky. For most of the results presented here, we restrict the sky location to the known position of SSS17a/AT 2017gfo as determined by electromagnetic observations [12]. In every case, we use a prior on the distance which assumes a homogeneous rate density in the nearby Universe, with no cosmological corrections applied; in other words, the distance prior grows with the square of the luminosity distance. Meanwhile, we use EM observations to reweight our distance posteriors when investigating the inclination of the binary in Sec. III A, and we use the measured redshift factor to the host galaxy NGC 4993 in order to infer source-frame masses from detector-frame masses in Sec. III B. For the angle $\cos\theta_{JN} = \hat{\mathbf{J}} \cdot \hat{\mathbf{N}}$, defined for the total angular momentum $\mathbf{J}$ and the line of sight $\mathbf{N}$, we assume a prior distribution uniform in $\cos\theta_{JN}$ [104]. To improve the convergence rate of the stochastic samplers, the analyses with the nonprecessing waveform models implement a likelihood function where the phase at coalescence is analytically marginalized out [45].

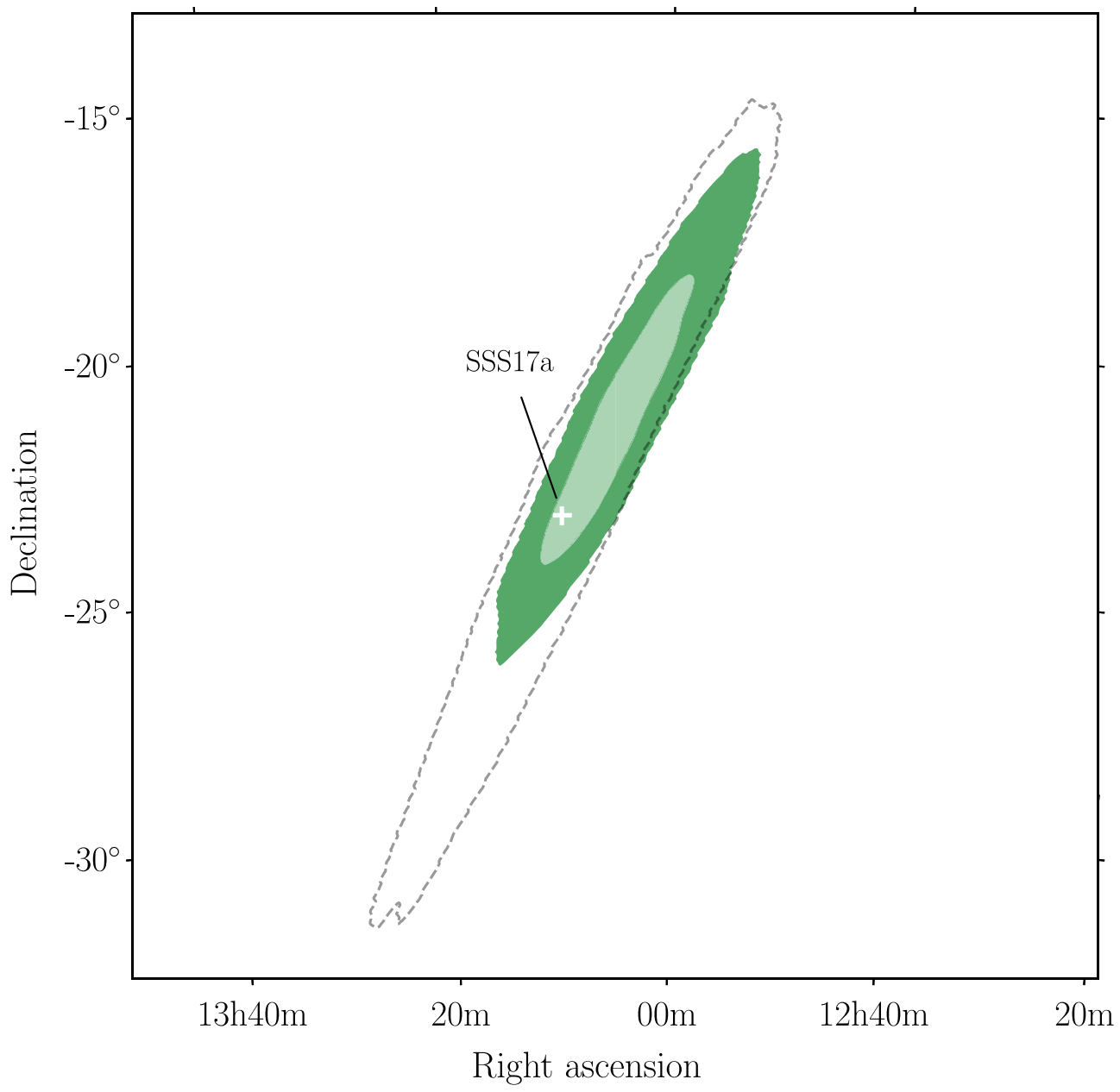

FIG. 3. The improved localization of GW170817, with the location of the associated counterpart SSS17a/AT 2017gfo. The darker and lighter green shaded regions correspond to 50% and 90% credible regions, respectively, and the gray dashed line encloses the previously derived 90% credible region presented in Ref. [3].

## III. PROPERTIES INFERRED FROM INSPIRAL AND MERGER

### A. Localization

For most of the analyses in this work, we assume *a priori* that the source of GW170817 is in NGC 4993. However, the improved calibration of Virgo data enables better localization of the source of GW170817 from GW data alone. To demonstrate the improved localization, we use results from the updated TaylorF2 analysis (the choice of model does not meaningfully affect localization [105]), shown in Fig. 3. We find a reduction in the 90% localization region from 28 $\mathrm{deg}^2$ [3] to 16 $\mathrm{deg}^2$. This improved localization is still consistent with the associated counterpart SSS17a/AT 2017gfo (see Fig. 3).

For the remainder of this work, we incorporate our knowledge of the location of the event.

While fixing the position of the event to the known location within NGC 4993, we infer the luminosity distance from the GW data alone. Using the PhenomPNRT waveform model, we find that the luminosity distance is $D_L = 41^{+6}_{-12}$ Mpc in the high-spin case and $D_L = 39^{+7}_{-14}$ Mpc in the low-spin case. Combining this distance information with the redshift associated with the Hubble flow at NGC 4993, we measure the Hubble parameter as in Ref. [106]. We find that $H_0 = 70^{+13}_{-7}\ \mathrm{km\,s^{-1}\,Mpc^{-1}}$ (we use maximum *a posteriori* and 68.3% credible interval for only $H_0$ in this work) in the high-spin case and $H_0 = 70^{+19}_{-8}\ \mathrm{km\,s^{-1}\,Mpc^{-1}}$ in the low-spin case; both measurements are within the uncertainties seen in Extended Data Table I and Extended Data Fig. 2 of Ref. [106]. As noted in Refs. [106,107], when only measuring one polarization of GW radiation from a binary merger, in the absence of strong precession there is a degeneracy between distance and inclination of the binary. When using GW170817 to measure the Hubble constant, this degeneracy is the main source of uncertainty. The slightly stronger constraints on $H_0$ in the high-spin case arise because, under that prior, our weak constraint on precession (see Sec. III C) helps to rule out binary inclinations that are closer to edge-on (i.e., $\theta_{JN} = 90$ deg) and where precession effects would be measurable, and hence increases the lower bound on the luminosity distance. Meanwhile, the upper bound on the luminosity distance is achieved with face-off (i.e., $\theta_{JN} = 180$ deg) binary inclinations and is nearly the same for both high-spin and low-spin cases.

This same weak constraint on precession leads to a tighter constraint on the inclination angle in the high-spin case when using the precessing signal model PhenomPNRT, $\theta_{JN} = 152^{+21}_{-27}$ deg, as compared to the low-spin case. The inclination measurement in the low-spin case, $\theta_{JN} = 146^{+25}_{-27}$ deg, agrees with the inferred values for both the high- and low-spin cases of our three waveform models that treat only aligned spins (see Table IV in Appendix A). This agreement gives further evidence that it is the absence of strong precession effects in the signal, which can only occur in the high-spin case of the precessing model, that leads to tighter constraints on $\theta_{JN}$. These tighter constraints are absent for systems restricted to the lower spins expected from Galactic NS binaries.

Conversely, EM measurements of the distance to the host galaxy can be used to reduce the effect of this degeneracy, improving constraints on the luminosity distance of the binary and its inclination, which may be useful for constraining emission mechanisms. Figure 4 compares our posterior estimates for distance and inclination with no *a priori* assumptions regarding the distance to the binary (i.e., using a uniform-in-volume prior) to the improved constraints from an EM-informed prior for the distance to the binary. For the EM-informed results, we have reweighted the posterior distribution to use a prior in distance following a normal distribution with mean 40.7 Mpc and standard deviation 2.36 Mpc [108]. This leads to improved measurements of the inclination angle $\theta_{JN} = 151^{+15}_{-11}$ deg (low spin) and $\theta_{JN} = 153^{+15}_{-11}$ deg (high spin). This measurement is consistent for both the high-spin and low-spin cases since the EM measurements constrain the source of GW170817 to higher luminosity distances and correspondingly more face-off inclination values. They are also consistent with the limits reported in previous studies using afterglow measurements [109] and combined

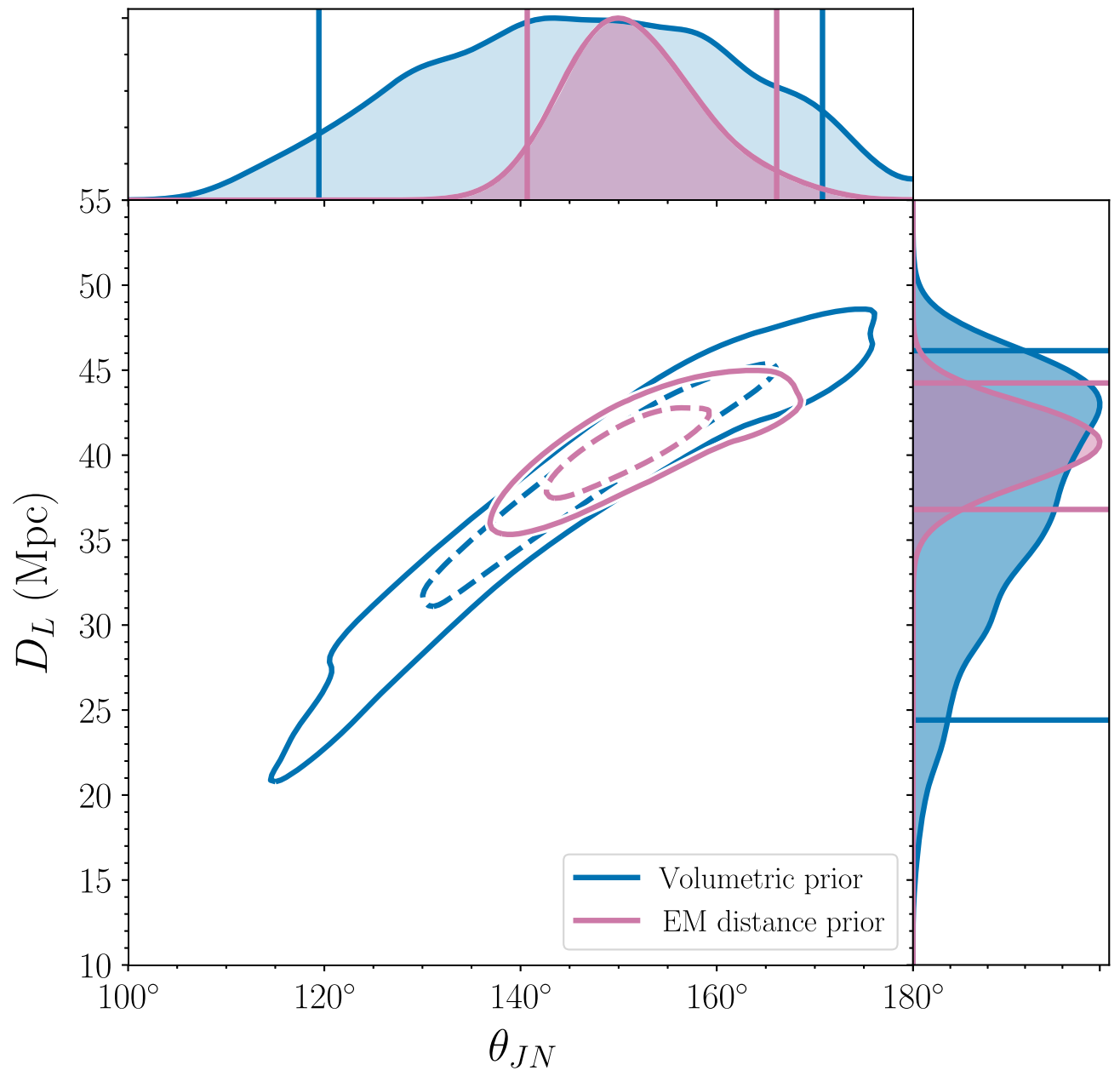

FIG. 4. Marginalized posteriors for the binary inclination ($\theta_{JN}$) and luminosity distance ($D_L$) using a uniform-in-volume prior (blue) and EM-constrained luminosity distance prior (purple) [108]. The dashed and solid contours enclose the 50% and 90% credible regions, respectively. Both analyses use a low-spin prior and make use of the known location of SSS17a. The 1D marginal distributions have been renormalized to have equal maxima to facilitate comparison, and the vertical and horizontal lines mark 90% credible intervals.

GW and EM constraints [108,110,111] to infer the inclination of the binary.

### B. Masses

Owing to its low mass, most of the SNR for GW170817 comes from the inspiral phase, while the merger and postmerger phases happen at frequencies above 1 kHz, where LIGO and Virgo are less sensitive (Fig. 1). This case is different than the BBH systems detected so far, e.g., GW150914 [112–115] or GW170814 [53]. The inspiral phase evolution of a compact binary coalescence can be written as a PN expansion, a power series in $v/c$, where $v$ is the characteristic velocity within the system [65]. The intrinsic parameters on which the system depends enter the expansion at different PN orders. Generally speaking, parameters that enter at lower orders have a large impact on the phase evolution and are thus easier to measure using the inspiral portion of the signal.

The chirp mass $\mathcal{M}$ enters the phase evolution at the lowest order; thus, we expect it to be the best constrained among the source parameters [32,61,93,94]. The mass ratio $q$, and consequently the component masses, are instead harder to measure due to two main factors: (1) They are higher-order corrections in the phase evolution, and (2) the mass ratio is partially degenerate with the component of the spins aligned with the orbital angular momentum [93,94,116], as discussed further below.

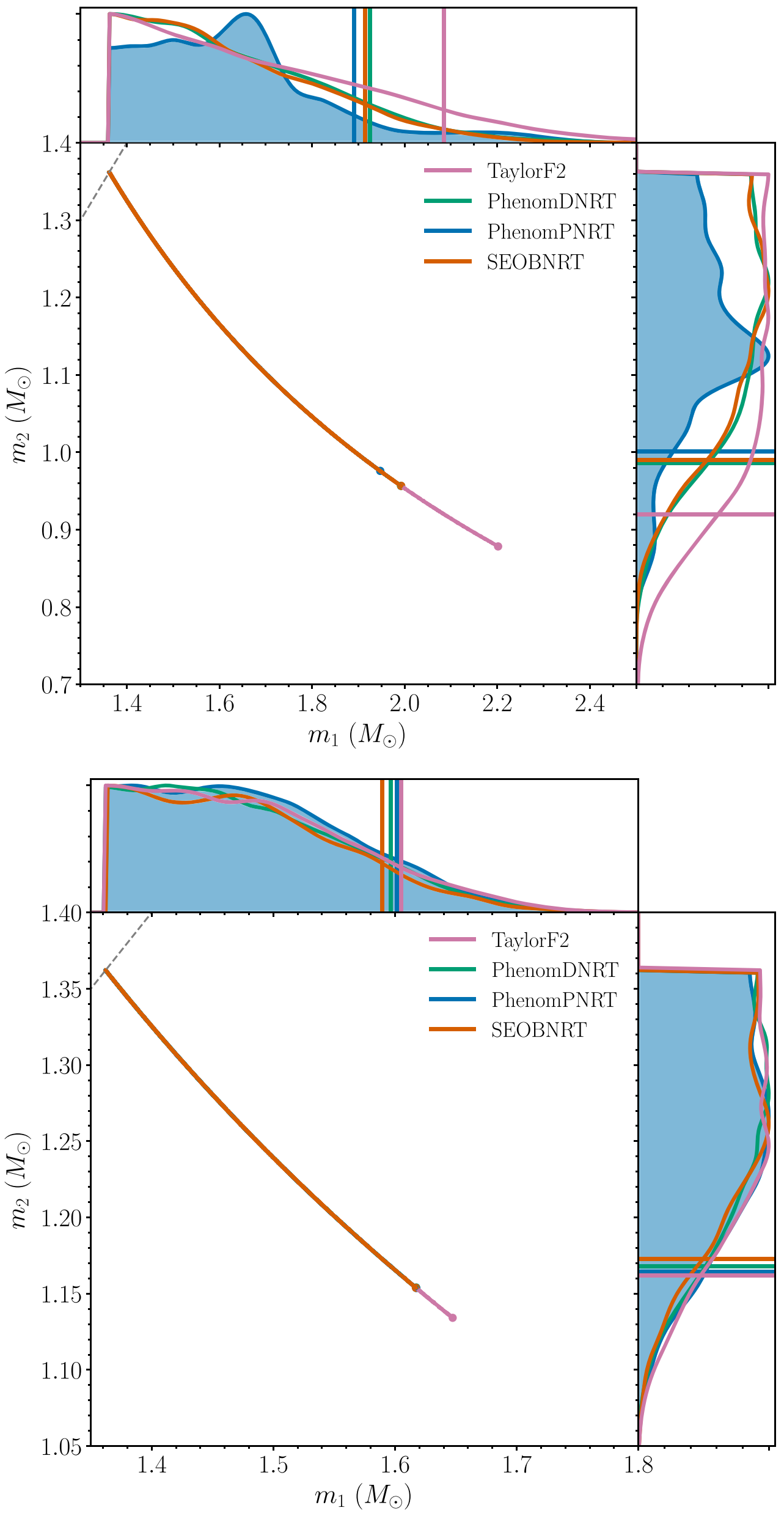

FIG. 5. The 90% credible regions for component masses using the four waveform models for the high-spin prior (top panel) and low-spin prior (bottom panel). The true thickness of the contour, determined by the uncertainty in the chirp mass, is too small to show. The points mark the edge of the 90% credible regions. The 1D marginal distributions have been renormalized to have equal maxima, and the vertical and horizontal lines give the 90% upper and lower limits on $m_1$ and $m_2$, respectively.

In Fig. 5, we show one-sided 90% credible intervals of the joint posterior distribution of the two component masses in the source frame. We obtain $m_1 \in (1.36, 1.89)$ $M_\odot$ and $m_2 \in (1.00, 1.36)$ $M_\odot$ in the high-spin case, and tighter constraints of $m_1 \in (1.36, 1.60)$ $M_\odot$ and $m_2 \in (1.16, 1.36)$ $M_\odot$ in the low-spin case. These estimates are consistent with, and generally more precise than, those presented in Ref. [3]. The inferred masses for the components are also broadly consistent with the known masses of Galactic neutron stars observed in BNS systems (see, e.g., Ref. [117]).

TABLE II. Properties for GW170817 inferred using the PhenomPNRT waveform model. All properties are source properties except for the detector-frame chirp mass $\mathcal{M}^{\rm det} = \mathcal{M}(1+z)$. Errors quoted as $x^{+z}_{-y}$ represent the median, 5% lower limit, and 95% upper limit. Errors quoted as $(x, y)$ are one-sided 90% lower or upper limits, and they are used when one side is bounded by a prior. For the masses, $m_1$ is bounded from below and $m_2$ is bounded from above by the equal-mass line. The mass ratio is bounded by $q \leq 1$. For the tidal parameter $\tilde{\Lambda}$, we quote results using a constant (flat) prior in $\tilde{\Lambda}$. In the high-spin case, we quote a 90% upper limit for $\tilde{\Lambda}$, while in the low-spin case, we report both the symmetric 90% credible interval and the 90% highest posterior density (HPD) interval, which is the smallest interval that contains 90% of the probability.

| | Low-spin prior ($\chi \leq 0.05$) | High-spin prior ($\chi \leq 0.89$) |
|---|---|---|
| Binary inclination $\theta_{JN}$ | $146^{+25}_{-27}$ deg | $152^{+21}_{-27}$ deg |
| Binary inclination $\theta_{JN}$ using EM distance constraint [108] | $151^{+15}_{-11}$ deg | $153^{+15}_{-11}$ deg |
| Detector-frame chirp mass $\mathcal{M}^{\rm det}$ | $1.1975^{+0.0001}_{-0.0001}$ $M_\odot$ | $1.1976^{+0.0004}_{-0.0002}$ $M_\odot$ |
| Chirp mass $\mathcal{M}$ | $1.186^{+0.001}_{-0.001}$ $M_\odot$ | $1.186^{+0.001}_{-0.001}$ $M_\odot$ |
| Primary mass $m_1$ | $(1.36, 1.60)$ $M_\odot$ | $(1.36, 1.89)$ $M_\odot$ |
| Secondary mass $m_2$ | $(1.16, 1.36)$ $M_\odot$ | $(1.00, 1.36)$ $M_\odot$ |
| Total mass $m$ | $2.73^{+0.04}_{-0.01}$ $M_\odot$ | $2.77^{+0.22}_{-0.05}$ $M_\odot$ |
| Mass ratio $q$ | $(0.73, 1.00)$ | $(0.53, 1.00)$ |
| Effective spin $\chi_{\rm eff}$ | $0.00^{+0.02}_{-0.01}$ | $0.02^{+0.08}_{-0.02}$ |
| Primary dimensionless spin $\chi_1$ | $(0.00, 0.04)$ | $(0.00, 0.50)$ |
| Secondary dimensionless spin $\chi_2$ | $(0.00, 0.04)$ | $(0.00, 0.61)$ |
| Tidal deformability $\tilde{\Lambda}$ with flat prior | $300^{+500}_{-190}$(symmetric)/$300^{+420}_{-230}$(HPD) | $(0, 630)$ |

As expected, the detector-frame chirp mass is measured with much higher precision, with $\mathcal{M}^{\rm det} = 1.1976^{+0.0004}_{-0.0002}$ $M_\odot$ (high spin) and $\mathcal{M}^{\rm det} = 1.1975^{+0.0001}_{-0.0001}$ $M_\odot$ (low spin). These uncertainties are decreased by nearly a factor of 2 as compared to the value reported in Ref. [3] for the detector-frame chirp mass, while the median remains consistent with the 90% credible intervals previously reported. The main source of uncertainty in the source-frame chirp mass comes from the unknown velocity of the source: The line-of-sight velocity dispersion $\sigma_v = 170$ km s$^{-1}$ of NGC 4993 reported in Ref. [96] translates into an uncertainty on the geocentric redshift of the source $z = 0.0099 \pm 0.0009$, and thereby onto the chirp mass. This uncertainty dominates over the statistical uncertainty in $\mathcal{M}$ and over the subpercent level uncertainty in the redshift measurement of NGC 4993 reported in Ref. [97]. The use of the velocity dispersion to estimate the uncertainty in the radial velocity of the source is consistent with the impact of the second supernova on the center-of-mass velocity of the progenitor of GW170817 being relatively small [96,118], especially given that the probable delay time of GW170817 is much longer than the dynamical time of its host galaxy. Both sources of uncertainty are incorporated into the values reported in Table II, which still correspond to a subpercent level of precision on the measurement of $\mathcal{M}$. This method of determining $\mathcal{M}$ from the detector-frame chirp mass differs from the original method used in Ref. [3], and the resulting median value of $\mathcal{M}$ lies at the edge of the 90% credible interval reported there, with uncertainties reduced by a factor of 2 or more. The fact that chirp mass is estimated much better than the individual masses is the reason why, in Fig. 5, the two-dimensional posteriors are so narrow in one direction [119]. Meanwhile, the unknown velocity of the progenitor of GW170817 impacts the component masses at a subpercent level and is neglected in the bounds reported above and in Table II.

### C. Spins

The spins of compact objects directly impact the phasing and amplitude of the GW signal through gravitomagnetic interactions (e.g., Refs. [120–122]) and through additional contributions to the mass- and current-multipole moments, which are the sources of GWs (e.g., Ref. [65]). This result allows for the measurement of the spins of the compact objects from their GW emission. The spins produce two qualitatively different effects on the waveform.

First, the components of spins along the orbital angular momentum **L** have the effect of slowing down or speeding up the overall rate of inspiral, for aligned-spin components and anti-aligned spin components, respectively [123]. The most important combination of spin components along **L** is a mass-weighted combination called the effective spin, $\chi_{\rm eff}$ [124–126], defined as

$$\chi_{\rm eff} = \frac{m_1 \chi_{1z} + m_2 \chi_{2z}}{m_1 + m_2}. \quad (3)$$

This combination contributes to the gravitational-wave phase evolution at 1.5PN order, together with $\mathcal{M}$, $q$, and an additional spin degree of freedom [93,127]. This leads to a degeneracy among these quantities, especially between $q$ and $\chi_{\rm eff}$, which complicates the measurement of both of these parameters from the GW phase. The remaining aligned-spin degree of freedom at 1.5PN order is more important for systems with lower mass ratios [128], while the perpendicular components of the spins first contribute to the phasing at 2PN order [61,127,129].

Second, the components of the spins perpendicular to the instantaneous direction of **L** precess due to spin-orbit and spin-spin interactions. This leads to the precession of the orbital plane itself in order to approximately conserve the direction of the total angular momentum, which modulates the GW phasing and signal amplitude measured by a fixed observer [104]. One benefit of considering the effective spin $\chi_{\rm eff}$ is that it is approximately conserved throughout inspiral, even as the other components of spins undergo complicated precessional dynamics [130].

The precession-induced modulations of the GW amplitude and phase occur on timescales that span many orbital periods. They are most measurable for systems with large spin components perpendicular to **L**, for systems with smaller mass ratios $q$, and for systems viewed close to edge-on [131,132], where precession of the orbital plane strongly modulates the observed signal [104]. Precession effects are commonly quantified by an effective spin-precession parameter $\chi_p$, which is defined as [133]

$$\chi_p = \max\left(\chi_{1\perp}, \frac{3+4q}{4+3q} q \chi_{2\perp}\right), \quad (4)$$

where $\chi_{i\perp}$ are the magnitudes of the components of the dimensionless spins which are perpendicular to **L**. When considering precessing binaries, we must specify a reference frequency at which spin-related quantities such as $\chi_p$ and the individual spins are extracted. For the precessing waveform PhenomPNRT used in this work, we use 100 Hz.

As discussed previously, we use two choices for priors on component spins, a prior that allows for high spins ($\chi \leq 0.89$) and one that restricts to lower spin magnitudes ($\chi \leq 0.05$). The choice of prior has a strong impact on our spin inferences, which in turn influences the inferred component masses through the $q$–$\chi_{\rm eff}$ degeneracy.

Figure 6 shows the marginalized posterior probability distributions for $\chi_{\rm eff}$ from the four waveform models, along with the high-spin and low-spin priors. For the high-spin case, we find that negative values of $\chi_{\rm eff}$ are mostly excluded for all of the models, although small negative $\chi_{\rm eff}$ and negligible values are still allowed. Large values of $\chi_{\rm eff}$ are also excluded, and the 90% credible interval for PhenomPNRT is $\chi_{\rm eff} \in (-0.00, 0.10)$. The uncertainty in $\chi_{\rm eff}$ is reduced by nearly a factor of 2 as compared with the more conservative constraint $\chi_{\rm eff} \in (-0.01, 0.17)$ reported in Ref. [3] for this prior, and it remains consistent with negligibly small spins. For the low-spin prior, the constraints on negative values of $\chi_{\rm eff}$ are nearly identical, but in this case, the upper end of the $\chi_{\rm eff}$ marginal posterior is shaped by the prior distribution. The 90% credible interval in the low-spin case for PhenomPNRT is $\chi_{\rm eff} \in (-0.01, 0.02)$, which is the same range as reported in Ref. [3] for the low-spin case.

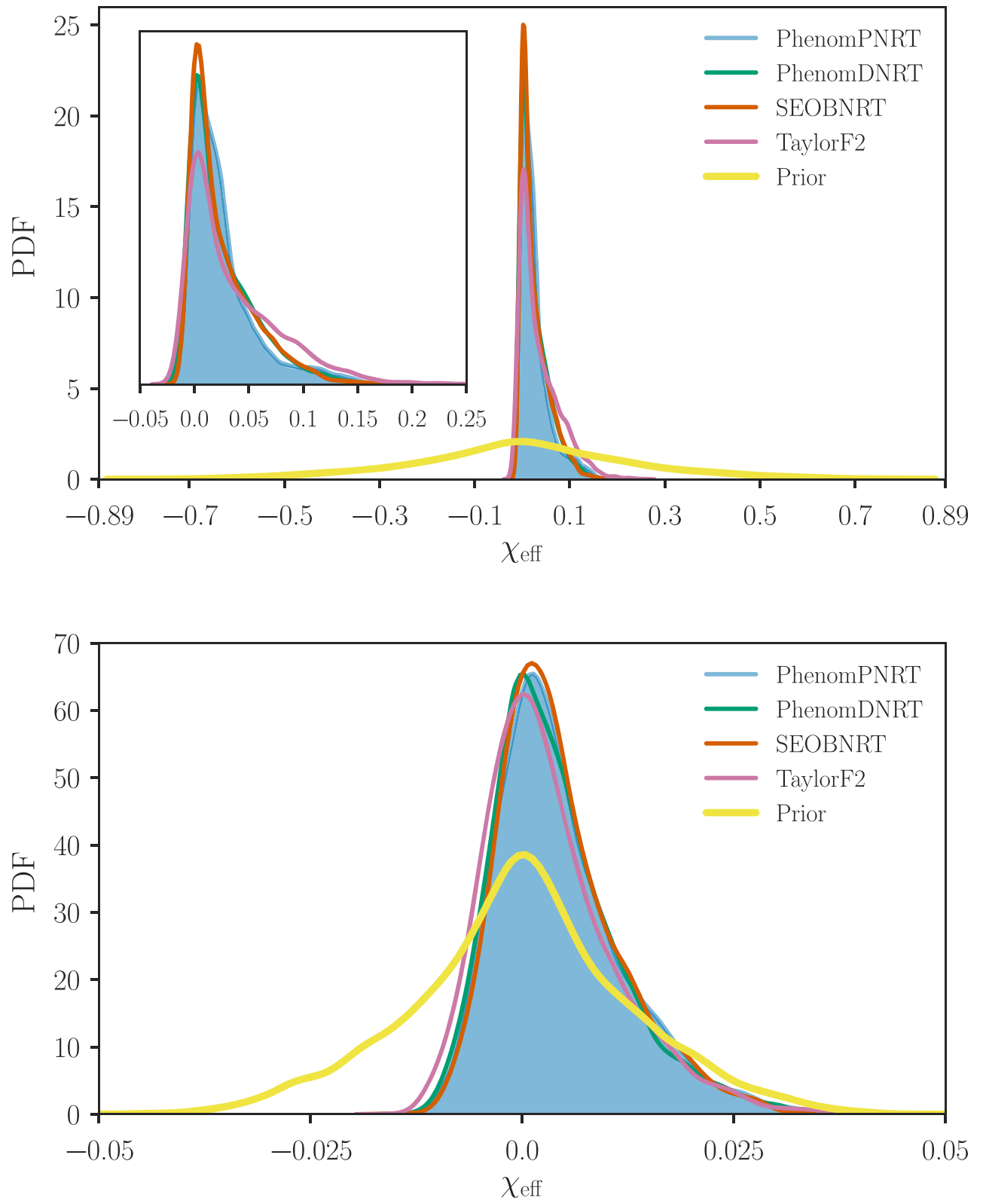

FIG. 6. Posterior PDF for the effective spin parameter $\chi_{\rm eff}$ using the high-spin prior (top panel) and low-spin prior (bottom panel). The four waveform models used are TaylorF2, PhenomDNRT, PhenomPNRT, and SEOBNRT.

Figure 7 shows two-dimensional marginalized posteriors for $q$ and $\chi_{\rm eff}$ for PhenomPNRT, illustrating the degeneracy between these parameters. The two-dimensional posterior distributions are truncated at the boundary $q = 1$, and when combined with the degeneracy, this causes a positive skew in the marginalized $\chi_{\rm eff}$ posteriors, as seen in Fig. 6 [134]. Compared to the high-spin priors, the low-spin prior on $\chi_{\rm eff}$ cuts off smaller values of $q$, favoring nearly equal-mass systems.

While all of the models provide constraints on the effective spin, only the PhenomPNRT model provides constraints on the spin precession of the binary. The top panel of Fig. 8 shows the inferred component spin magnitudes and orientations for the high-spin case. In the high-spin case, Fig. 8 shows that we rule out large spin components aligned or anti-aligned with $\mathbf{L}$, but the constraints on in-plane spin components are weaker. As such, we can only rule out large values for the effective precession parameter $\chi_p$, as seen in the bottom panel of Fig 8, with the upper 90th percentile at 0.53. Nevertheless, in this case, we can place bounds on the magnitudes of the component spins; we find that the 90% upper bounds are $\chi_1 \leq 0.50$ and $\chi_2 \leq 0.61$, still well above the range of spins inferred for Galactic binary neutron stars.

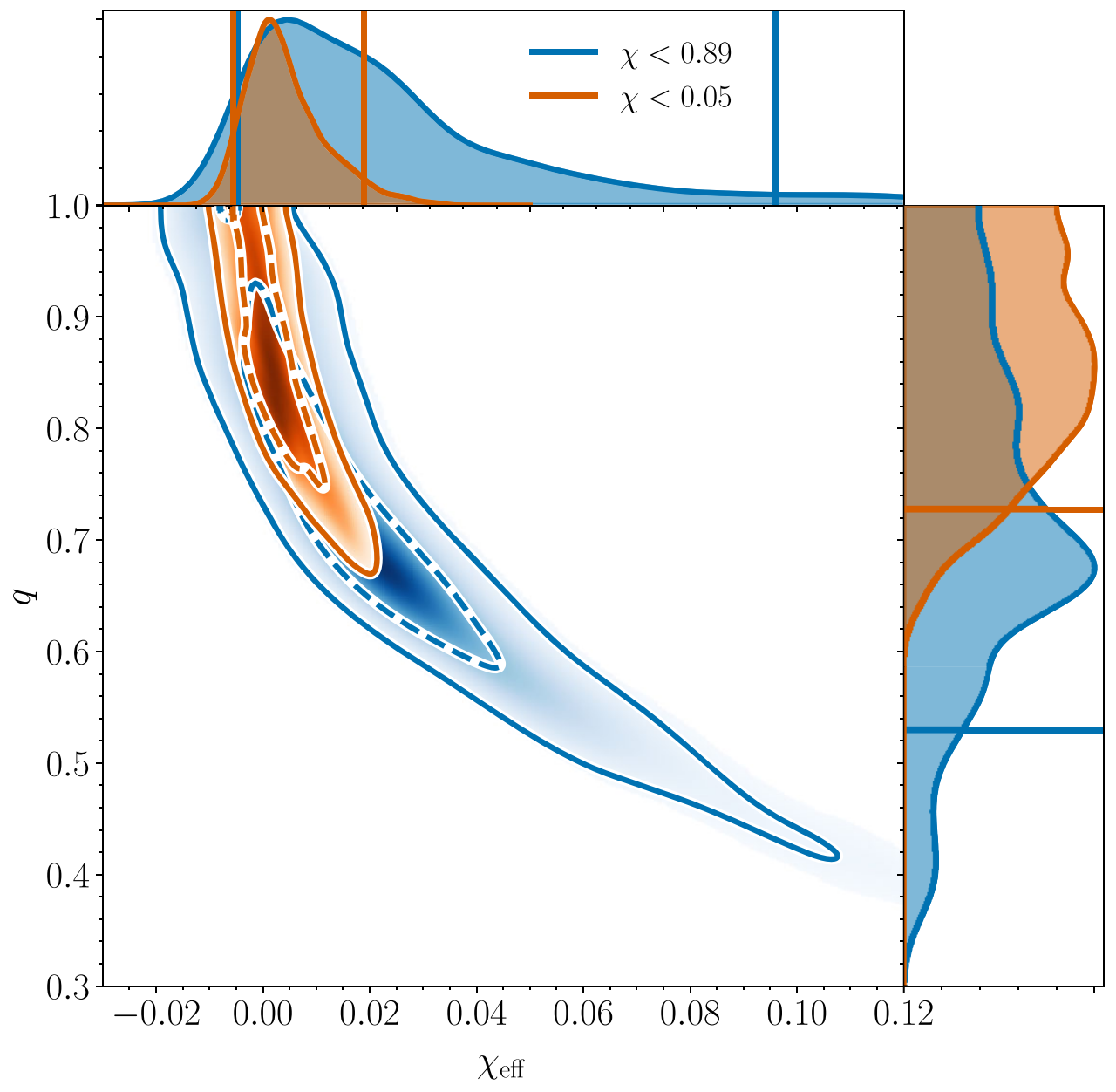

FIG. 7. Marginalized two-dimensional posteriors for the effective spin $\chi_{\mathrm{eff}}$ and mass ratio $q$ using the PhenomPNRT model for the high-spin prior (blue) and low-spin prior (orange). The 50% (dashed line) and 90% (solid line) credible regions are shown for the joint posterior. The 90% credible interval for $\chi_{\mathrm{eff}}$ is shown by vertical lines, and the 90% lower limit for $q$ is shown by horizontal lines. The 1D marginal distributions have been renormalized to have equal maxima.

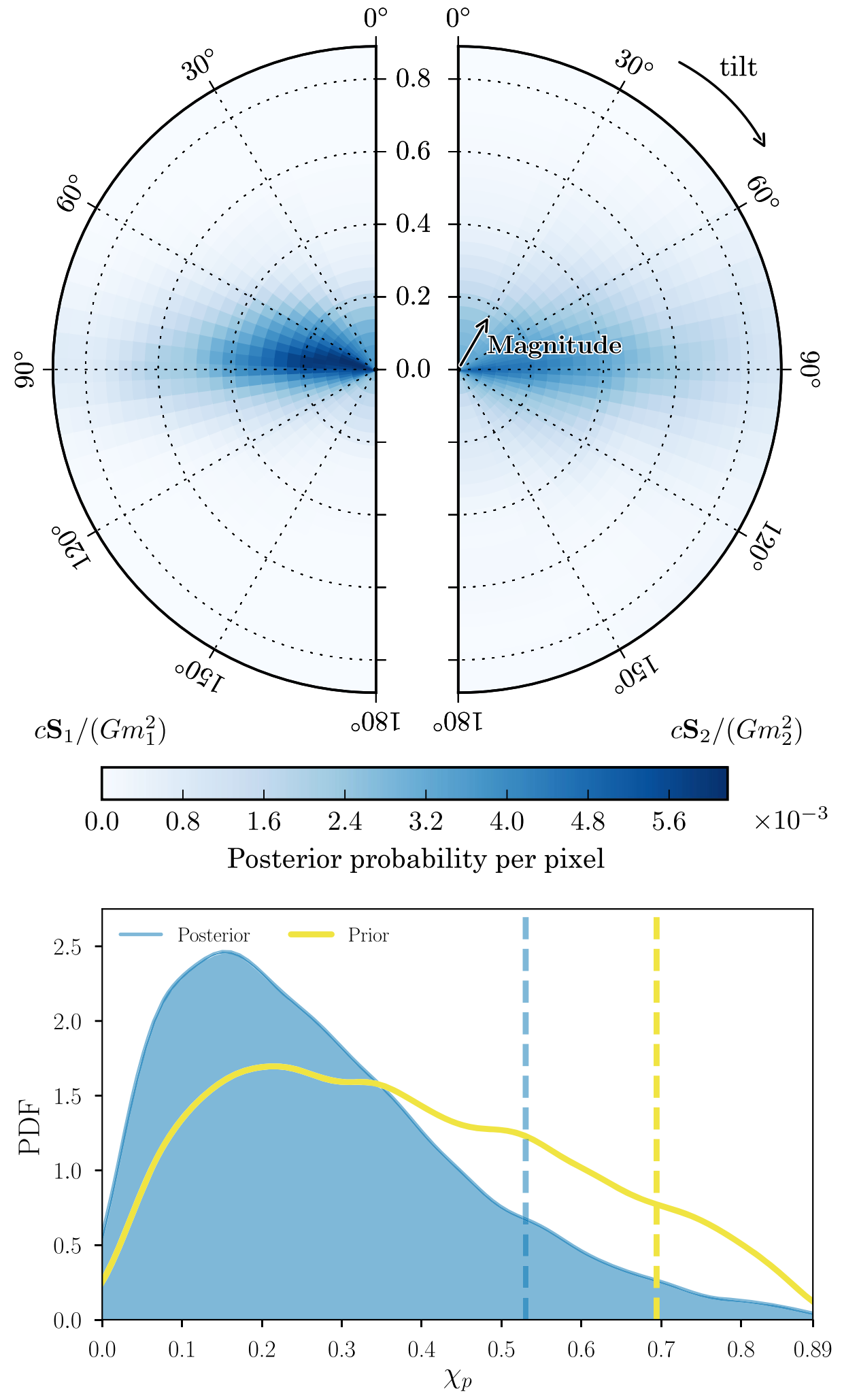

FIG. 8. Top panel: Inferred spin parameters using the PhenomPNRT model in the high-spin case, where the dimensionless component spin magnitudes $\chi < 0.89$. Plotted are the probability densities for the dimensionless spin components $\boldsymbol{\chi}_1$ and $\boldsymbol{\chi}_2$ relative to the orbital angular momentum $\mathbf{L}$, plotted at the reference gravitational-wave frequency of $f = 100$ Hz. A tilt angle of 0° indicates alignment with $\mathbf{L}$. Each pixel has equal prior probability. Bottom panel: The posterior for the precession parameter $\chi_p$, plotted together with its prior distribution, also plotted at the reference frequency of $f = 100$ Hz. The vertical lines represent the 90th percentile for each distribution.

Figure 9 shows the same quantities as Fig. 8 using the low-spin prior. In this case, we primarily constrain the spins to lie in or above the orbital plane at the reference frequency. This is consistent with the inferences on $\chi_{\mathrm{eff}}$, which rule out large negative values of $\chi_{\mathrm{eff}}$ but whose upper bounds are controlled by the prior distribution. Meanwhile, for $\chi_p$, the upper 90th percentile is at 0.04, which is nearly unchanged between the prior and posterior distributions. The inability to place strong constraints on precession is consistent with an analysis reported in Ref. [3] using a precessing model that neglects tidal effects [78].

### D. Tidal parameters

In the post-Newtonian formalism, matter effects for nonspinning objects first enter the waveform phase at 5PN order through the tidally induced quadrupolar ($\ell = 2$) deformation [135]. The amount of deformation is described by the dimensionless tidal deformability of each NS, defined by $\Lambda = (2/3)k_2[(c^2/G)(R/m)]^5$, where $k_2$ is the dimensionless $\ell = 2$ Love number and $R$ is the NS radius. These quantities depend on the NS mass $m$ and EOS. For spinning NSs, matter effects also enter at 2PN due to the spin-induced quadrupole moment as discussed in Sec. II C, and of the models considered here, only PhenomPNRT implements this effect.

We show marginalized posteriors for the tidal parameters $\Lambda_1$ and $\Lambda_2$ in Fig. 10 for the four waveform models. For TaylorF2, the results in this work are in general agreement with the values reported in Ref. [3], which also used the TaylorF2 model. However, here we use a lower starting frequency of 23 Hz instead of 30 Hz, resulting in upper bounds on $\Lambda_1$ and $\Lambda_2$ that are about 10% (for the high-spin prior) and about 20% (for the low-spin prior) smaller than in Ref. [3]. This improvement occurs because, although most of the tidal effects occur above several hundred Hz as shown in Fig. 2, the tidal parameters still have a weak correlation with the other parameters. Using more low-frequency information improves the measurement of the other parameters and thus decreases correlated uncertainties in the tidal parameters.

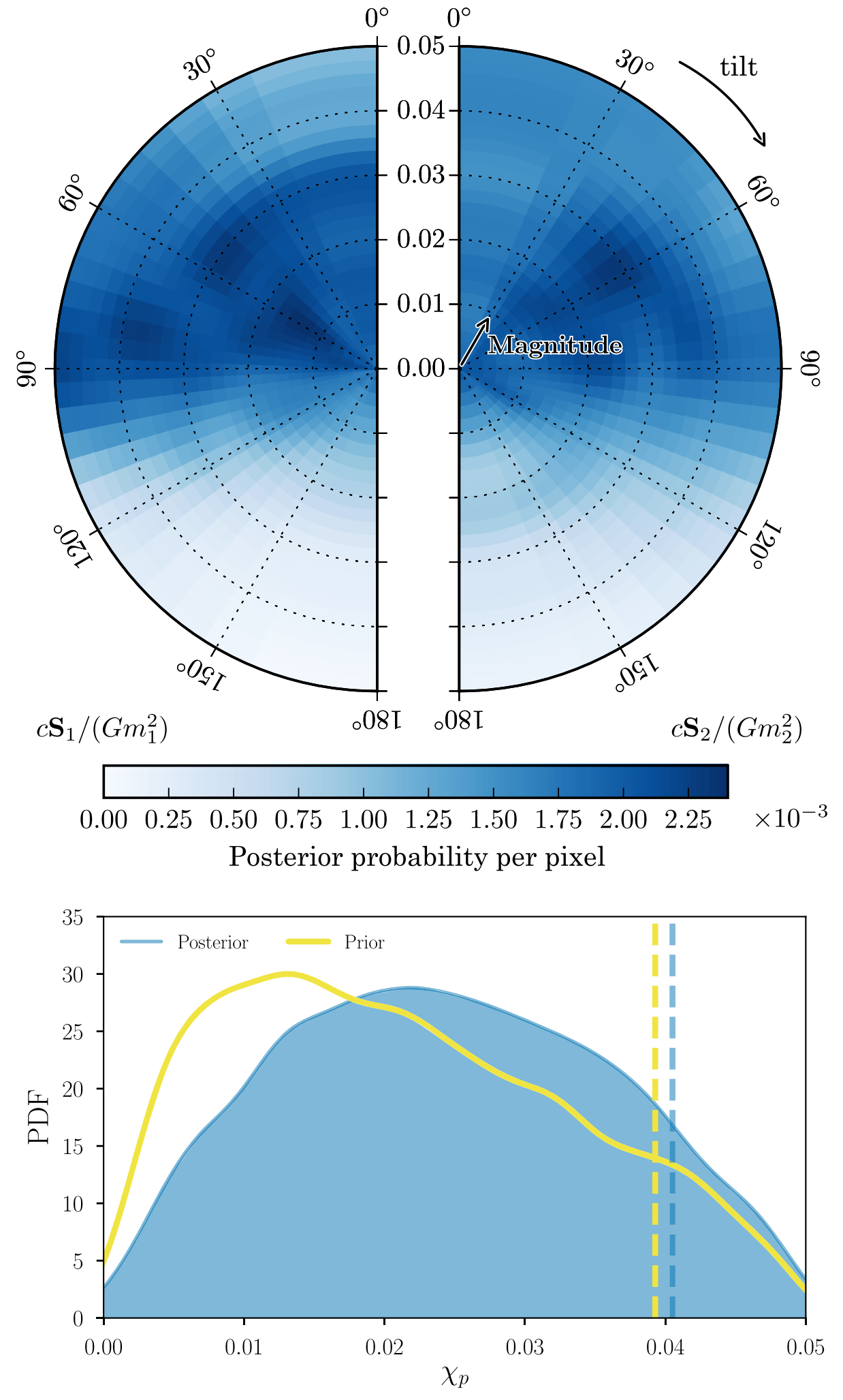

FIG. 9. Inferred spin parameters using the PhenomPNRT model as in Fig. 8, but in the low-spin case where the dimensionless component spin magnitudes $\chi < 0.05$. The posterior probability densities for the dimensionless spin components and for $\chi_p$ are plotted at the reference gravitational-wave frequency of $f = 100$ Hz.

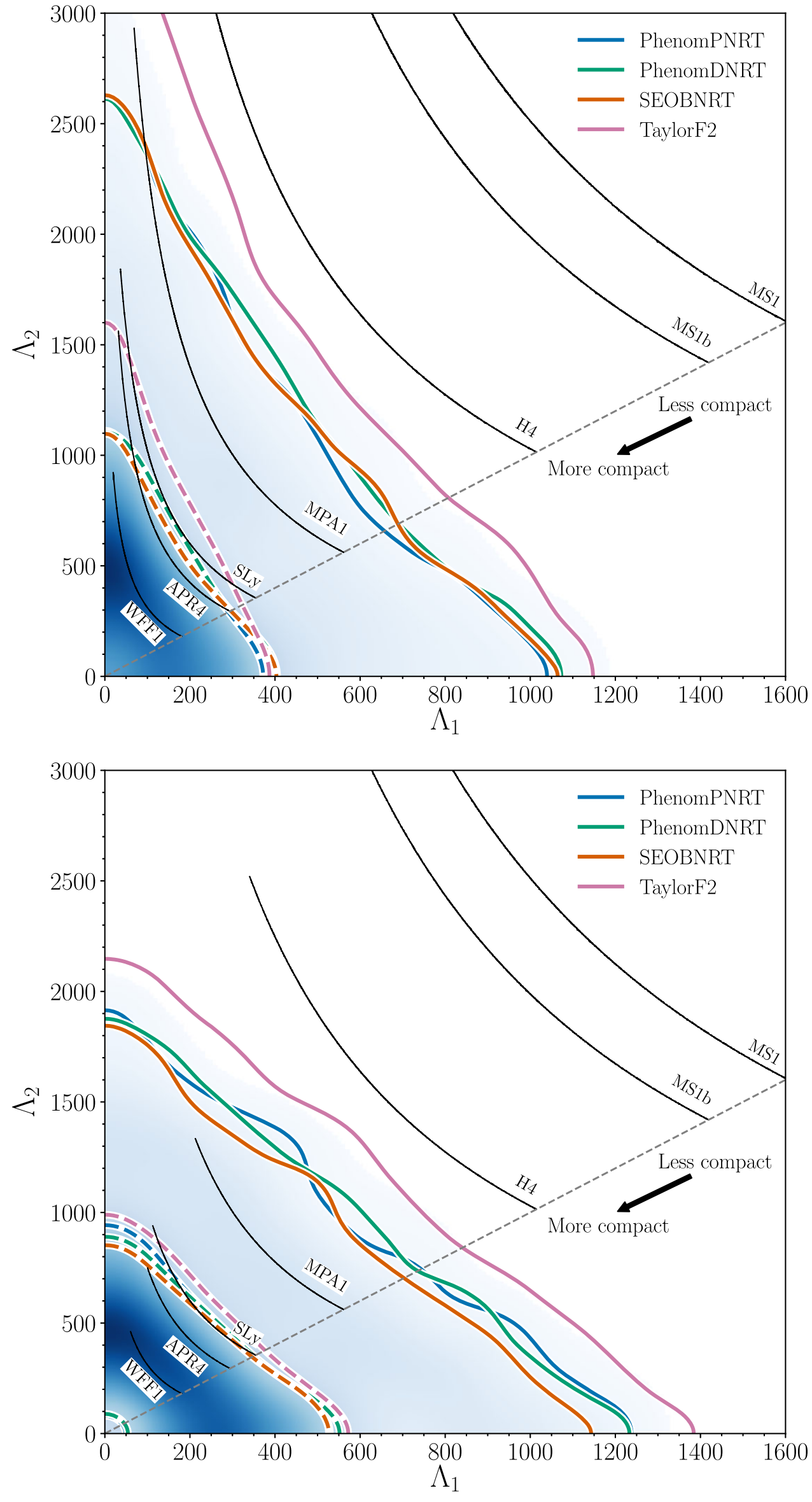

FIG. 10. PDFs for the tidal deformability parameters $\Lambda_1$ and $\Lambda_2$ using the high-spin (top panel) and low-spin (bottom panel) priors. The blue shading is the PDF for the precessing waveform PhenomPNRT. The 50% (dashed lines) and 90% (solid lines) credible regions are shown for the four waveform models. The seven black curves are the tidal parameters for the seven representative EOS models using the masses estimated with the PhenomPNRT model, ending at the $\Lambda_1 = \Lambda_2$ boundary.

The three waveform models that use the same NRTidal prescription produce nearly identical 90% upper limits that are about 10% smaller than those of TaylorF2. The reason for this result is that the tidal effect for these models is larger than for TaylorF2, as shown in Fig. 2, so the tidal parameters that best fit the data will be smaller in order to compensate. Including precession and the spin-induced quadrupole moment in the PhenomPNRT model does not noticeably change the results for the tidal parameters compared to the other two models with the NRTidal prescription. Overall, as already found in Ref. [3] the NRTidal models have 90% upper limits that are about 20%–30% lower than the TaylorF2 results presented.

For reference, we also show $\Lambda_1$–$\Lambda_2$ contours for a representative subset of theoretical EOS models that span the range of plausible tidal parameters using piecewise-polytrope fits from Refs. [136,137]. The values of $\Lambda_1$ and $\Lambda_2$ are calculated using the samples for the source-frame masses $m_1$ and $m_2$ contained in the 90% credible region for PhenomPNRT. The widths of these bands are determined by the small uncertainty in chirp mass. The lengths of these bands are determined by the uncertainty in mass ratio. Most of their support is near the $\Lambda_1 = \Lambda_2$ line corresponding to the equal-mass case and ends at the 90% lower limit for the mass ratio. The predicted values of the tidal parameters for the EOSs MS1, MS1b, and H4 lie well outside of the 90% credible region for both the low-spin and high-spin priors, and for all waveform models. This can be compared to Fig. 5 of Ref. [3], where H4 was still marginally consistent with the 90% credible region.

The leading tidal contribution to the GW phase evolution is a mass-weighted linear combination of the two tidal parameters $\tilde{\Lambda}$ [138]. It first appears at 5PN order and is defined such that $\tilde{\Lambda} = \Lambda_1 = \Lambda_2$ when $m_1 = m_2$:

$$\tilde{\Lambda} = \frac{16}{13} \frac{(m_1 + 12m_2)m_1^4\Lambda_1 + (m_2 + 12m_1)m_2^4\Lambda_2}{(m_1 + m_2)^5}. \quad (5)$$

In Fig. 11, we show marginalized posteriors of $\tilde{\Lambda}$ for the two spin priors and four waveform models. Because there is only one combination of the component tidal deformabilities that gives $\tilde{\Lambda} = 0$, namely, $\Lambda_1 = \Lambda_2 = 0$, when using flat priors in $\Lambda_1$ and $\Lambda_2$, the prior distribution for $\tilde{\Lambda}$ falls to zero as $\tilde{\Lambda} \to 0$. This means that the posterior for $\tilde{\Lambda}$ must also fall to zero as $\tilde{\Lambda} \to 0$. To avoid the misinterpretation that there is no evidence for $\tilde{\Lambda} = 0$, we reweight the posterior for $\tilde{\Lambda}$ by dividing by the prior used, effectively imposing a flat prior in $\tilde{\Lambda}$. In practice, this is done by dividing a histogram of the posterior by a histogram of the prior. The resulting histogram is then resampled and smoothed with kernel density estimation. We have verified the validity of the reweighting procedure by comparing the results to runs where we fix $\Lambda_2 = 0$ and use a flat prior in $\tilde{\Lambda}$. This differs from the reweighting procedure only in the small, next-to-leading-order tidal effect.

After reweighting, there is still some support at $\tilde{\Lambda} = 0$. For the high-spin prior, we can only place a 90% upper limit on the tidal parameter, shown in Fig. 11 and listed in Tables II and IV. For the TaylorF2 model, this 90% upper limit can be directly compared to the value reported in Ref. [3]. We note, however, that due to a bookkeeping error, the value reported in Ref. [3] should have been 800 instead of 700. Our improved value of 730 is about 10% less than this corrected value. As with the $\Lambda_1$–$\Lambda_2$ posterior (Fig. 10), the three models with the NRTidal prescription predict 90% upper limits that are consistent with each other and less than the TaylorF2 results by about 10%. For the low-spin prior, we can now place a two-sided 90% HPD credible interval on $\tilde{\Lambda}$ that does not contain $\tilde{\Lambda} = 0$. This 90% HPD interval is the smallest interval that contains 90% of the probability.

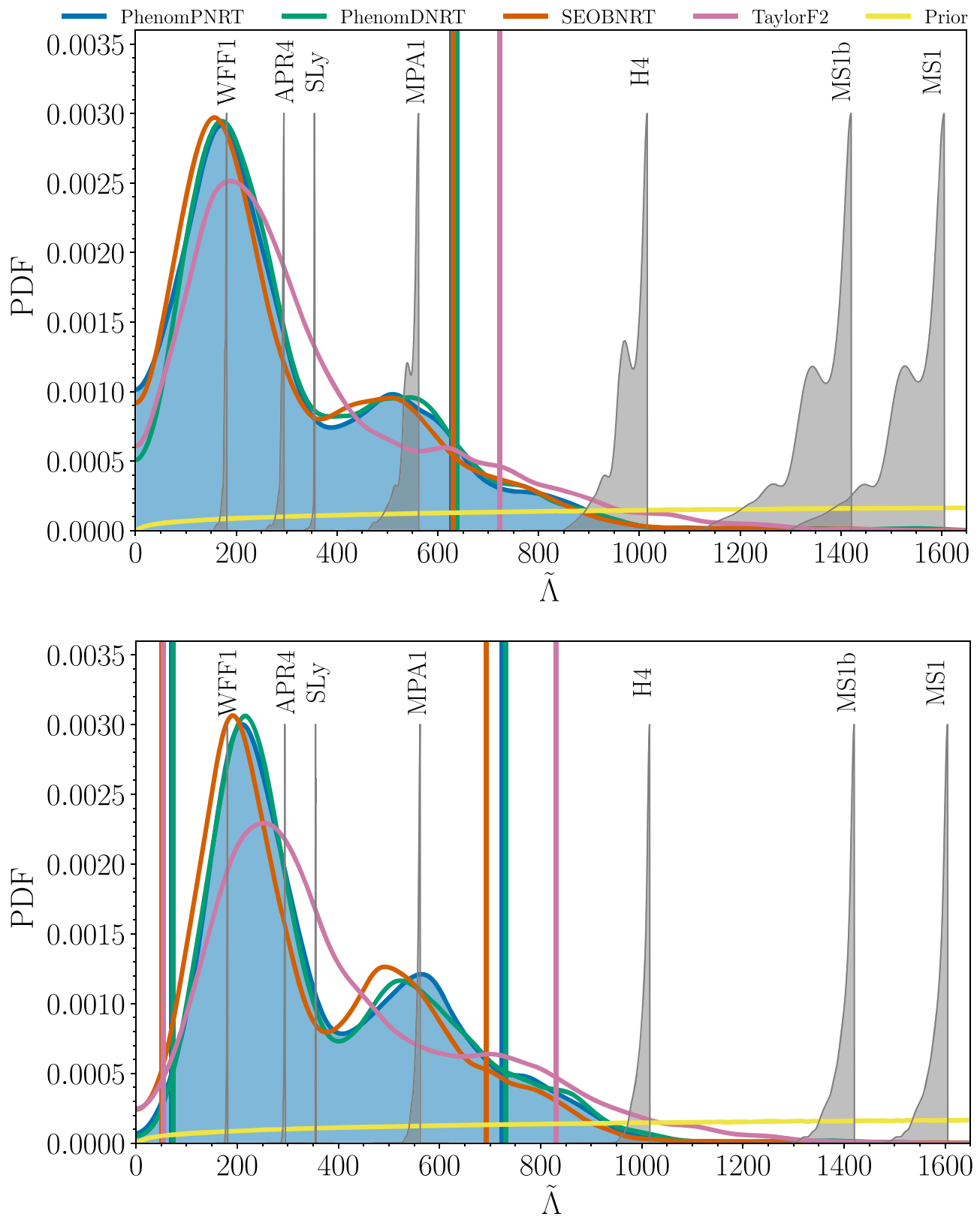

FIG. 11. PDFs of the combined tidal parameter $\tilde{\Lambda}$ for the high-spin (top panel) and low-spin (bottom panel) priors. Unlike in Fig. 6, the PDFs have been reweighted by dividing by the original prior for $\tilde{\Lambda}$ (also shown). The 90% HPD credible intervals are represented by vertical lines for each of the four waveform models: TaylorF2, PhenomDNRT, SEOBNRT, and PhenomPNRT. For the high-spin prior, the lower limit on the credible interval is $\tilde{\Lambda} = 0$. The seven gray PDFs are those for the seven representative EOSs using the masses estimated with the PhenomPNRT model. Their normalization constants have been rescaled to fit in the figure. For these EOSs, a 1.36 $M_\odot$ NS has a radius of 10.4 km (WFF1), 11.3 km (APR4), 11.7 km (SLy), 12.4 km (MPA1), 14.0 km (H4), 14.5 km (MS1b), and 14.9 km (MS1).

The PDFs for the NRTidal waveform models are bimodal. The secondary peak's origin is the subject of further investigation, but it may result from a specific noise realization, as similar results have been seen with injected waveforms with simulated Gaussian noise (see Fig. 4 of Ref. [138]).

In Fig. 11, we also show posteriors of $\tilde{\Lambda}$ (gray PDFs) predicted by the same (EOSs) as in Fig. 10, evaluated using the masses $m_1$ and $m_2$ sampled from the posterior. The sharp cutoff to the right of each EOS posterior corresponds to the equal-mass-ratio boundary. Again, as in Fig. 10, the EOSs MS1, MS1b, and H4 lie outside the 90% credible upper limit and are therefore disfavored.

The differences between the high-spin prior and low-spin prior can be better understood from the joint posterior for $\tilde{\Lambda}$ and the mass ratio $q$. Figure 12 shows these posteriors for the PhenomPNRT model without reweighting by the prior. For mass ratios near $q = 1$, the two posteriors are similar. However, the high-spin prior allows for a larger range of mass ratios, and for smaller values of $q$, there is more support for small values of $\tilde{\Lambda}$. If we restrict the mass ratio to $q \gtrsim 0.5$, or equivalently $m_2 \gtrsim 1\ M_\odot$, we find that there is less support for small values of $\tilde{\Lambda}$, and the two posteriors for $\tilde{\Lambda}$ are nearly identical.

To verify that we have reliably measured the tidal parameters, we supplement the four waveforms used in this paper with two time-domain EOB waveform models: SEOBNRv4T [81,139] and TEOBResumS [80]. SEOBNRv4T includes dynamical tides and the effects of the spin-induced quadrupole moment. TEOBResumS incorporates a gravitational-self-force resummed tidal potential and the spin-induced quadrupole moment. Both models are compatible with state-of-the-art BNS numerical simulations up to merger [83,140].

Unfortunately, these waveform models are too expensive to be used for parameter estimation with LALInference.

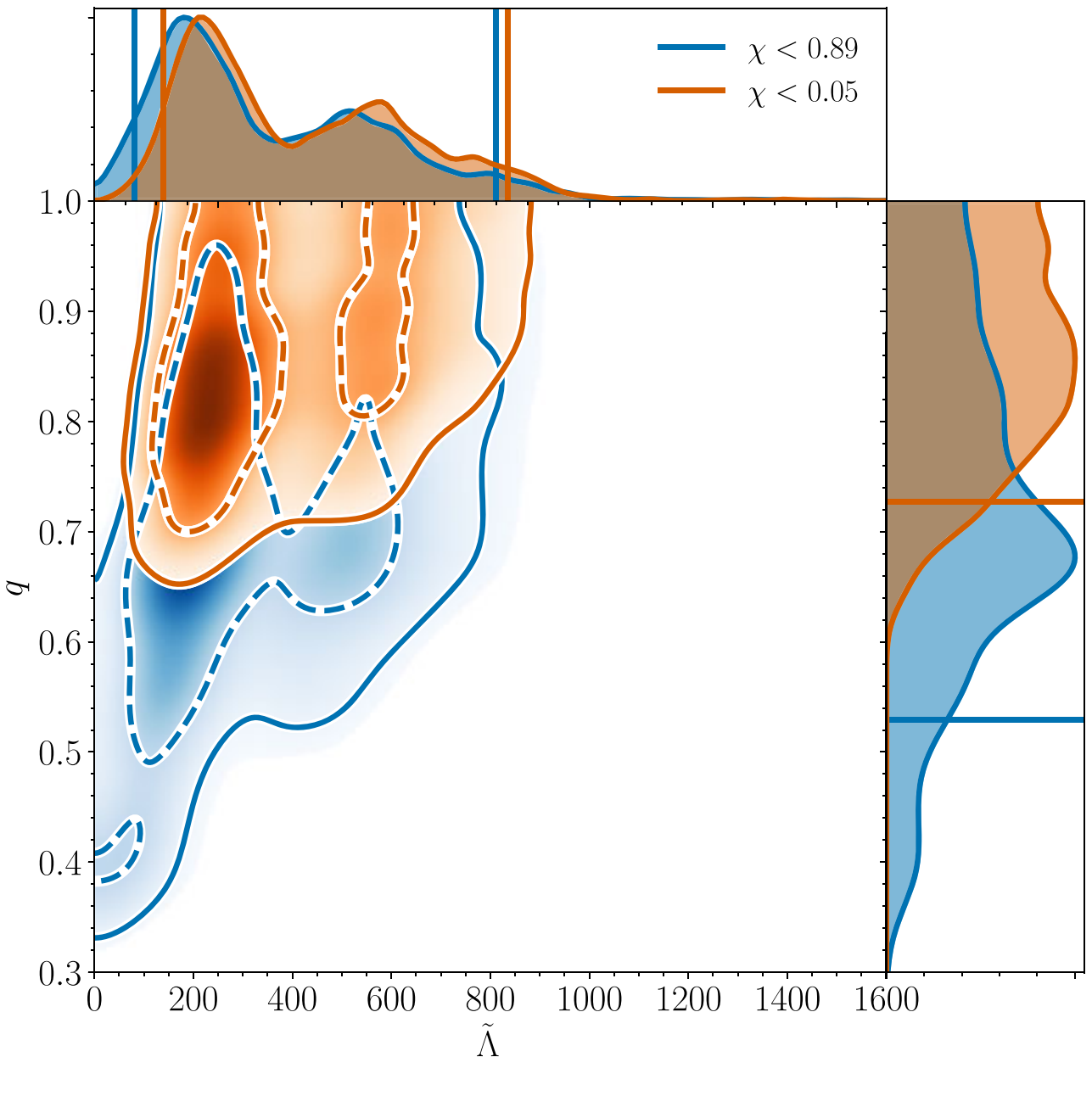

FIG. 12. PDFs for the tidal parameter $\tilde{\Lambda}$ and mass ratio $q$ using the PhenomPNRT model for the high-spin (blue) and low-spin (orange) priors. Unlike Fig. 11, the posterior is not reweighted by the prior, so the support that is seen at $\tilde{\Lambda} = 0$ is due to smoothing from the kernel density estimator (KDE) that approximates the distribution from the discrete samples. The 50% (dashed lines) and 90% (solid lines) credible regions are shown for the joint posterior. The 90% credible interval for $\tilde{\Lambda}$ is shown by vertical lines, and the 90% lower limit for $q$ is shown by horizontal lines.

We therefore use the parallelized but less validated parameter estimation code RapidPE [84,85]. This code uses a different procedure from the standard LALInference code for generating posterior samples and allows for parameter estimation with significantly more expensive waveform models. For each point in the intrinsic parameter space, RapidPE marginalizes over the extrinsic parameters with Monte Carlo integration. For aligned-spin models, the resulting six-dimensional intrinsic marginalized posterior is then adaptively sampled and fit with Gaussian process regression. Samples from this fitted posterior are then drawn using a Markov-chain Monte Carlo algorithm.

We perform runs with RapidPE using the low-spin prior for three waveform models. The first run uses the PhenomDNRT waveform for a direct comparison with the LALInference result. The 90% highest posterior density credible interval for $\tilde{\Lambda}$ is shifted downward from (70,730) using LALInference to (20,690) using RapidPE. Although these differences are not negligible, they are still smaller than the differences between different waveform models. The main difference, however, is that $\tilde{\Lambda}$ has a bimodal structure using LALInference that is not seen with RapidPE. There are several possible reasons for this difference. One possibility is over-smoothing from the Gaussian process regression fit used in RapidPE. Another possibility is the difference in data processing when evaluating the likelihood functions for the two codes. In addition, RapidPE does not marginalize over detector calibration uncertainties. However, comparisons using LALInference with and without calibration error marginalization show that this cannot account for the differences between LALInference and RapidPE. Unfortunately, we have not been able to resolve the differences in the shape of the posterior. Given its extensive previous use and testing, we use LALInference for our main results, and we only use RapidPE for exploratory studies, leaving detailed comparisons to future work. For the two EOB waveforms, the 90% highest posterior density credible interval for $\tilde{\Lambda}$ is (0,560) for SEOBNRv4T and (10,690) for TEOBResumS. For SEOBNRv4T, the posterior for $\tilde{\Lambda}$ has a peak away from $\tilde{\Lambda} = 0$, and the lower bound of $\tilde{\Lambda} = 0$ is not simply due to the prior bound. In fact, the value of the posterior distribution is the same at both the upper and lower limits of the 90% credible interval, indicating that the peak is resolved.

Recently, De *et al.* performed an independent analysis of the GW data to measure the tidal parameters [141]. Their results are broadly consistent with those presented here but are made under the assumption that the two merging NSs have the same EOS. They assume that the two NSs have identical radii and that the tidal deformabilities of the individual stars are related by the approximate relation $\Lambda_1 = q^6 \Lambda_2$, whereas we allow the tidal parameters to vary independently. A more direct comparison of the results is made in our companion paper, where we assume a common EOS using approximate universal relations as well as directly sampling a parametrized EOS [42,101–103,142,143].

## IV. LIMITS ON POSTMERGER SIGNAL

Having used the inspiral phase of the GW signal to constrain the properties of the component bodies, we now place limits on the signal content after the two stars merged to make inferences about the remnant object. The outcome of a BNS coalescence depends on the progenitor masses and the NS EOS. Soft EOSs and large masses result in the prompt formation of a black hole immediately after the merger [144]. Stiffer EOS and lower masses result in the formation of a stable or quasistable NS remnant [145,146]. A hypermassive NS, whose mass exceeds the maximum mass of a uniformly rotating star but is supported by differential rotation and possibly thermal gradients [145], will survive for $\lesssim 1$s, after which time the NS collapses into a black hole [147,148]. A supramassive star, whose mass is lower but still exceeds the threshold for nonrotating NSs, will spin down on longer timescales before forming a black hole [149]. Finally, extremely stiff EOSs and low masses will result in a stable NS.

We use the BayesWave algorithm [39] to form frequency-dependent upper limits on the strain amplitude and radiated energy by following the approach described in Ref. [150]. BayesWave models GWs as a superposition of an arbitrary number of elliptically polarized Morlet-Gabor wavelets. This signal model has been found to be capable of accurate waveform reconstruction for a variety of signal morphologies, including short-duration postmerger signals [150]. The priors of this analysis are expressed in terms of the individual wavelet parameters and on the SNR of each wavelet. Consequently, the priors on the signal amplitude and waveform morphology are derived from the individual wavelet priors, rather than being directly specified. The priors on the wavelet quality factor and phase are flat in (0,200) and $(0, 2\pi)$, respectively. The priors on the central frequency and time are determined by the analysis duration and bandwidth described below, while the amplitude prior is determined through the SNR of each wavelet and discussed in more detail in Ref. [39].

We use the analysis described in Ref. [150] to estimate an upper bound on the amplitude of a putative GW signal assumed to be present but at insufficiently high SNR to generate a statistically significant detection candidate. We use coincident data from the two LIGO detectors and from GEO600 [40], which has comparable sensitivity to Virgo at high frequency. Indeed, the sky location of GW170817 is particularly favorable for the GEO600 antenna response so that any high-frequency signal component observed by GEO600 will have a SNR greater than or equal to that expected in Virgo. During this period, the Virgo data above 2 kHz suffer from an abundance of spectral lines and transient noise and, therefore, are not included in this analysis. It should also be noted that GEO600 was not in science mode due to investigations into a degraded squeezer phase error point signal leading to a reduced level of squeezing. At the time of the event, the investigations were passive observations. Otherwise, GEO600 was in nominal running condition. The calibration of the LIGO detectors is more uncertain above 2 kHz than at lower frequencies, but it is still within 8% in amplitude and 4 deg in phase [54]. The GEO600 calibration uncertainty is estimated to be within 15% in amplitude and 15 deg in phase in the 1–4 kHz band. GEO600 was not used in a previous search for high-frequency GW emission due to an insufficient characterization of data quality and analysis tuning, which would have been required for accurate background estimation [41]. The analysis reported in this work, by contrast, is a Bayesian characterization of an underlying signal, and it involves only the 1 s of data around the coalescence time of the merger, which relaxes the data quality requirements somewhat. Furthermore, the analysis configuration has been chosen based on studies of the expected signal (i.e., Ref. [150]), and it is not optimized to eliminate statistical outliers in a background distribution.

We use a 1-s segment of data centered around the time of coalescence, and we restrict the analysis to waveforms whose peak amplitude lies within a 250-ms window at the center of the segment. This window is sufficient to account for statistical or systematic uncertainties in the time-of-coalescence measurement inferred from the inspiral signal, and the total length of segment used encompasses the duration of postmerger signals predicted by numerical simulations for hypermassive NSs that eventually collapse to black holes. The analysis is performed over the 1024–4096-Hz band, which is sufficient to contain the full postmerger spectrum.

We determine the relative evidence for two models: that the on-source data are described by Gaussian noise only, or by Gaussian noise plus a GW signal as described in Refs. [151,152]. We find that the Gaussian noise model is strongly preferred, with a Bayes factor (evidence ratio) of 256.79 over the signal model. This result is consistent with both prompt collapse to a BH and with a postmerger signal that is too weak to be measurable with our current sensitivity. We further characterize the absence of a detectable signal by forming 90% credible upper limits on three measures of signal strength: (i) the network SNR, evaluated over 1–4 kHz, (ii) the strain amplitude spectral density (ASD), and (iii) the spectral energy density (SED).

We compute the 90% credible upper limit on the network SNR directly using the reconstructed waveform posterior. We exclude signal power in our analysis band with $\rho_{\rm net} > 6.7$ at the 90% level. The top panel of Fig. 13 reports the upper limits and expectations for the strain ASD induced in the LIGO-Hanford instrument. These limits are formed directly from the posterior probability distribution for the reconstructed waveform in the 1 s of data around the merger time measured from the premerger observations using a coherent analysis of data from all three detectors. The noise ASDs for each instrument are shown for comparison. As one would expect in the absence of a signal, the upper limits on strain amplitude approximately follow the shape of the noise spectrum. We also overlay a small set of spectra obtained from simulations of BNS mergers using different EOSs with extrinsic parameters (i.e., sky location, inclination, and distance) determined from the premerger analysis. Information about the simulations used is presented in Table III. Depending on the EOS, the analysis frequency band might contain significant contributions from the inspiral and merger phases of the coalescence. If the simulated waveforms are truncated at peak amplitude such that we only include the postmerger phase, the network SNR of each waveform is about 0.5.

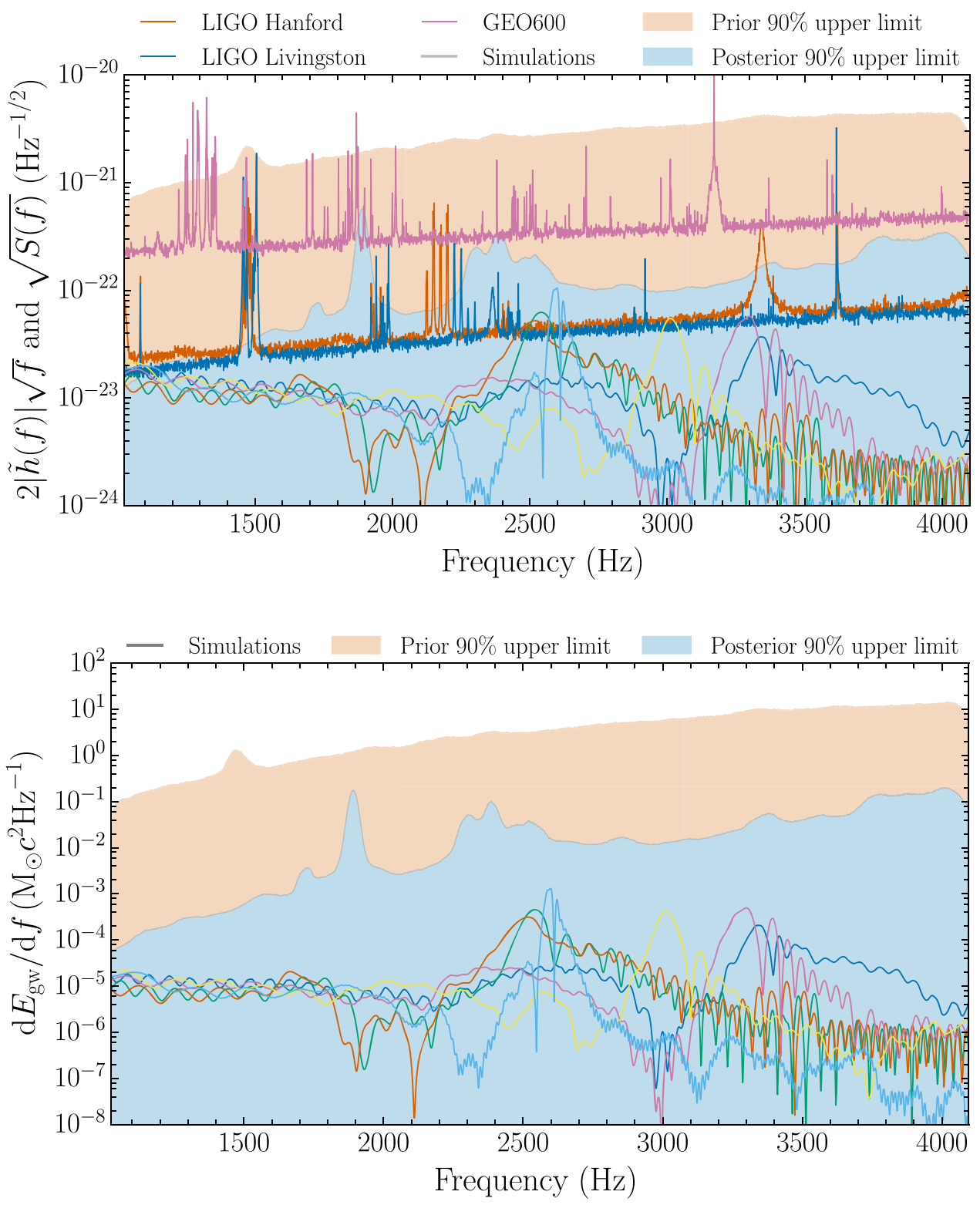

FIG. 13. The 90% credible upper limits on GW strain induced in the Hanford detector (top panel) and radiated energy (bottom panel). Both results are derived from a coherent analysis across the detector network. The noise ASDs for each instrument used in this analysis are shown for comparison top panel). Results from selected numerical simulations are also shown.

Finally, the peaklike structures evident in the posterior upper limit are due to low-significance instrumental artefacts and, particularly around 2.4 kHz, a nonstationary spectral line in LIGO-Livingston. The low end of the strain ASD posterior extends to include zero, consistent with the nondetection of a postmerger signal.

We apply a similar procedure to form a frequency-dependent 90% credible upper limit on GW energy (see Ref. [150] for details). The bottom panel of Fig. 13 shows the 90% credible upper limits on the SED. As with the reconstructed amplitude, the prior on the SED is imposed by the priors on individual wavelet parameters rather than any specific astrophysical argument. SEDs derived from BNS simulations in which the source is held at the distance, sky location, and orientation of GW170817 are shown for comparison with our upper limits. Our 90% credible upper limit is still too large to make any inference about the EOSs from this part of the signal. Instead, we characterize the sensitivity improvement required to begin to probe astrophysically interesting energy regimes by comparing the peaks of the simulated SEDs to the 90% credible upper limit on the energy radiated at that frequency.

We find that our upper limits on energy are 12–215 times larger than expectations based on our choice of EOS and simulations, shown in Table III. We therefore require amplitude sensitivity to improve by a factor of about 3.5–15 compared to our current results in order to probe realistic energy scales for an equivalent event. This should be regarded as a rather conservative estimate of the upgrade required before we can start probing the astrophysically interesting energy regime, as a number of improvements can increase the sensitivity of our analysis. The current methodology described in Ref. [150] is agnostic when it comes to the morphology of the postmerger signal. Additional information about the signal, such as its broadband structure or the finite extent of the postmerger peak, could increase the sensitivity of our analysis, making it easier to detect and characterize the postmerger signal.

As stated earlier, the analysis described here complements the previous, more generic, high-frequency search in Ref. [41]. The upper limits here are given by the 90% credible interval of the posterior probability distribution on the signal amplitude spectrum and its power spectral density. The analysis in Ref. [41], by contrast, reports the root-sum-squared amplitude that a number of numerical simulations would require in order that 50% of a population of those signals would produce a ranking statistic with false alarm probability of $10^{-4}$. Nonetheless, one can compare the amplitude sensitivity improvement required such that each analysis begins to probe astrophysically interesting energy scales. The two analyses share a subset of simulated signals: those with the H4, SLy, and SFHx EOSs reported in Table III. In Ref. [41], sensitivities are quoted in terms of the root-sum-squared amplitude, which scales with the square root of the gravitational-wave energy. The energy scales probed by Ref. [41] for the H4, SLy, and SFHx waveforms are, respectively, 169, 144, and 121 times higher than the values expected from merger simulations with extrinsic parameters of GW170817. In this analysis, the best energy upper limits for the same waveforms are 70, 64, and 31 times greater than the peak energies of those waveforms. With the caveat that we are free to compare our limits with the dominant postmerger frequency, the analysis reported here effectively probes a factor of about 2–4 smaller energies.

Sensitivity improvements may come from more stringent and accurate waveform models, serendipitously located sources, as well as improved instrumental high-frequency sensitivity. The Advanced LIGO design sensitivity, e.g., is expected to be 3 times better at high frequencies than what has been achieved to date [1,153], while squeezing is expected to improve the sensitivity by another factor of 2 [154]. Similarly, the high-frequency sensitivity of Virgo may see as much as a factor of about 40 improvement when design sensitivity is achieved [2,153]. The postmerger SNR of the simulated waveforms is about 6–8 times smaller than the SNR required for marginal reconstruction of the postmerger signal [150] depending on the EOS and its energy content [155]. A similar event observed with the full LIGO-Virgo network operating at design sensitivity would, therefore, offer an opportunity to probe an astrophysically interesting energy regime and may even provide an estimate of the dominant postmerger oscillation frequency and corresponding constraints on the NS EOS.

## V. CONCLUSIONS

This work provides the most constraining measurements of the source of GW170817 to date. Without imposing strong astrophysical priors on the masses or spins, we show that the GW data constrains the masses to the range expected for BNS systems and constrains spin components parallel to the orbital angular momentum to be small. The GW data, however, do not significantly constrain the spin components perpendicular to the orbital angular momentum. If there is significant spin, it must lie near the orbital plane of the binary. Imposing a prior on the distance to GW170817 from the known distance to the host NGC 4993 allows us to constrain the inclination angle of the binary, providing insight into the nature of gamma-ray bursts.

Our improved constraints on the tidal deformation of the binary components reduce the upper bounds on this deformation, further ruling out some of the stiffest equation-of-state models. In addition, we find evidence for finite-size effects by establishing a lower bound for the tidal deformation parameter $\tilde{\Lambda}$ when we restrict the spins to be within the ranges observed in Galactic binaries. However, when we allow for large component spins, we are still unable to rule out the possibility of no tidal deformation of the component stars, as would occur, e.g., in a surprisingly low-mass binary black hole merger. While the measured properties are consistent with what we expect for binary neutron star systems, we cannot definitively say from GW measurements alone that both components of the binary were indeed neutron stars.

Comparing results from four different waveform models provides assurance that systematic uncertainties are small compared to statistical uncertainties. Improved waveform models, as well as optimizations to the models and parameter estimation codes that allow them to be used, will further reduce systematic uncertainties. We have shown initial results with the RapidPE code and SEOBNRv4T and TEOBResumS waveform models, and found that the measured tidal parameters are consistent with the main results of the paper. Furthermore, updated instrumental calibration could improve constraints further. However, we do not expect these improvements to change the conclusions obtained here.

We have also placed new, morphology-agnostic bounds on the postmerger signal and argue that the Advanced LIGO-Virgo network at design sensitivity could have potentially reconstructed the postmerger signal.

There is still significant potential for improved constraints on GW170817 in the use of additional information in the priors for tidal deformation. In this work, we allow the component tidal parameters to vary independently, implicitly allowing each neutron star to have a different equation of state. One can require that the two neutron stars obey the same EOS through the use of binary universal relations [101–103] or a parametrized EOS [142,143]. This assumption also allows one to place bounds on the radii of the two neutron stars, and results are discussed in a companion paper [42].

TABLE III. Numerical simulations of the 1.35 $M_\odot$–1.35 $M_\odot$ binary neutron star mergers with different EOSs shown in Fig. 13. We report the value of the SED from each simulation at the peak frequency $f_{\mathrm{peak}}$ and our 90% credible upper limit on the SED at that frequency. The distance $D_L = 44.74$ Mpc and inclination $\theta_{JN} = 166.05$ deg are determined from the point of maximum posterior probability sampled from the PhenomPNRT model. Note that this is the maximum posterior probability sample drawn from the full posterior probability distribution, and it does not necessarily correspond to the maxima of the 1D and 2D marginal distributions shown in Fig. 4.

| EOS | Simulation | $f_{\mathrm{peak}}$ (Hz) | SED ($10^{-4}$ $M_\odot c^2$ Hz$^{-1}$) | SED$_{90}$ ($10^{-4}$ $M_\odot c^2$ Hz$^{-1}$) |
|---|---|---|---|---|
| APR4 [156] | [157] | 3342 | 2.1 | 450 |
| H4 [158] | [157] | 2541 | 4.5 | 320 |
| GNH3 [159] | [160] | 2522 | 3.1 | 380 |
| SLy [161] | [160] | 3299 | 5.0 | 320 |
| SFHx [162] | [163] | 3012 | 4.1 | 130 |
| DD2 [164,165] | [163] | 2598 | 13.1 | 160 |

Data associated with the figures in this article, including posterior samples generated using the PhenomPNRT model, can be found in Ref. [166].

## ACKNOWLEDGMENTS

The authors gratefully acknowledge the support of the United States National Science Foundation (NSF) for the construction and operation of the LIGO Laboratory and Advanced LIGO, as well as the Science and Technology Facilities Council (STFC) of the United Kingdom, the Max-Planck-Society (MPS), and the State of Niedersachsen/Germany for support of the construction of Advanced LIGO and construction and operation of the GEO600 detector. Additional support for Advanced LIGO was provided by the Australian Research Council. The authors gratefully acknowledge the Italian Istituto Nazionale di Fisica Nucleare (INFN), the French Centre National de la Recherche Scientifique (CNRS) and the Foundation for Fundamental Research on Matter supported by the Netherlands Organisation for Scientific Research, for the construction and operation of the Virgo detector and the creation and support of the EGO consortium. The authors also gratefully acknowledge research support from these agencies as well as by the Council of Scientific and Industrial Research of India; the Department of Science and Technology, India; the Science & Engineering Research Board (SERB), India; the Ministry of Human Resource Development, India; the Spanish Agencia Estatal de Investigación; the Vicepresidència i Conselleria d'Innovació; Recerca i Turisme and the Conselleria d'Educació i Universitat del Govern de les Illes Balears; the Conselleria d'Educació, Investigació, Cultura i Esport de la Generalitat Valenciana; the National Science Centre of Poland; the Swiss National Science Foundation (SNSF); the Russian Foundation for Basic Research; the Russian Science Foundation; the European Commission; the European Regional Development Funds (ERDF); the Royal Society; the Scottish Funding Council; the Scottish Universities Physics Alliance; the Hungarian Scientific Research Fund (OTKA); the Lyon Institute of Origins (LIO); the Paris Île-de-France Region; the National Research, Development and Innovation Office Hungary (NKFI); the National Research Foundation of Korea; Industry Canada and the Province of Ontario through the Ministry of Economic Development and Innovation; the Natural Science and Engineering Research Council Canada; the Canadian Institute for Advanced Research; the Brazilian Ministry of Science, Technology, Innovations, and Communications; the International Center for Theoretical Physics South American Institute for Fundamental Research (ICTP-SAIFR); the Research Grants Council of Hong Kong; the National Natural Science Foundation of China (NSFC); the Leverhulme Trust; the Research Corporation; the Ministry of Science and Technology (MOST), Taiwan; and the Kavli Foundation. The authors gratefully acknowledge the support of the NSF, STFC, Max-Planck-Society (MPS), INFN, CNRS, and the State of Niedersachsen/Germany for provision of computational resources. The GW strain data for this event are available at the LIGO Open Science Center [167]. This article has been assigned the document number LIGO-P1800061.

## APPENDIX A: SOURCE PROPERTIES FROM ADDITIONAL WAVEFORM MODELS

In this Appendix, we present additional results for the source properties of GW170817. Table IV presents the same inferred parameters quoted in Table II for the three additional waveform models TaylorF2, PhenomDNRT, and SEOBNRT. As expected from Figs. 5, 6, 10, and 11, the results among the four waveform models are largely consistent with each other.

One exception is the binary inclination angle $\theta_{JN}$: In the high-spin case, the precessing waveform PhenomPNRT achieves tighter bounds centered around a more face-off ($\theta_{JN} = 180$ deg) orientation than for the low-spin case. As discussed in Sec. III A, we attribute the tighter constraints on $\theta_{JN}$ to the fact that we disfavor configurations where strong precession effects would be observable; hence, we prefer values of $\theta_{JN}$ closer to face-off. Meanwhile, the other three waveform models only treat aligned spins, so the absence of strong precession does not help improve their inclination measurements. For all four waveforms, in the small-spin case, the spins are constrained to sufficiently small values so that there can be no strong precession effects; thus, again, the inclination measurements for the small-spin case are consistent with the aligned-spin measurements in the high-spin case. Finally, when we incorporate EM information about the distance to the source of GW170817, we eliminate the portion of the posteriors at closer distances and lower $\theta_{JN}$, achieving consistent inclination constraints across all cases.

The upper bounds on the spin magnitudes $\chi_1$ and $\chi_2$ are also lower for the waveforms that treat aligned spins only. This result is as expected, given that only the components $\chi_{i,z}$ contribute to the spin magnitudes for the aligned-spin runs, and these spin components are constrained by $\chi_{\rm eff}$ in the high-spin case and by our prior in the low-spin case. The differences between the remaining inferred parameters among the four waveforms give a sense of the possible size of systematic errors from our signal modeling, although PhenomPNRT includes the greatest number of relevant physical effects, as seen in Table I.

Table V presents the inferred intrinsic parameters of the binary as produced by RapidPE.

TABLE IV. Source properties for GW170817 using the additional waveform models TaylorF2, PhenomDNRT, and SEOBNRT. Conventions are the same as in Table II. The TaylorF2 results here can be directly compared with those from Ref. [3]. Note that the 90% upper limits for $\tilde{\Lambda}$ reported in Table I of Ref. [3] for TaylorF2 are incorrect (see Sec. III D). In Ref. [3], for the high-spin prior, it should be $\leq 800$ and not $\leq 700$, while for the low-spin prior, it should be $\leq 900$ and not $\leq 800$.

| High-spin prior, $\chi_i \leq 0.89$ | TaylorF2 | SEOBNRT | PhenomDNRT |
|---|---|---|---|
| Binary inclination $\theta_{JN}$ | $146^{+25}_{-28}$ deg | $146^{+24}_{-28}$ deg | $146^{+26}_{-28}$ deg |
| Binary inclination $\theta_{JN}$ using EM distance constraint [108] | $149^{+13}_{-10}$ deg | $152^{+14}_{-11}$ deg | $151^{+15}_{-10}$ deg |
| Detector-frame chirp mass $\mathcal{M}^{\rm det}$ | $1.1976^{+0.0004}_{-0.0002}$ $M_\odot$ | $1.1976^{+0.0003}_{-0.0002}$ $M_\odot$ | $1.1976^{+0.0003}_{-0.0002}$ $M_\odot$ |
| Chirp mass $\mathcal{M}$ | $1.186^{+0.001}_{-0.001}$ $M_\odot$ | $1.186^{+0.001}_{-0.001}$ $M_\odot$ | $1.186^{+0.001}_{-0.001}$ |
| Primary mass $m_1$ | $(1.36, 2.09)$ $M_\odot$ | $(1.36, 1.92)$ $M_\odot$ | $(1.36, 1.92)$ $M_\odot$ |
| Secondary mass $m_2$ | $(0.92, 1.36)$ $M_\odot$ | $(0.99, 1.36)$ $M_\odot$ | $(0.99, 1.36)$ $M_\odot$ |
| Total mass $m$ | $2.79^{+0.30}_{-0.06}$ $M_\odot$ | $2.76^{+0.20}_{-0.04}$ $M_\odot$ | $2.77^{+0.20}_{-0.04}$ $M_\odot$ |
| Mass ratio $q$ | (0.44, 1.00) | (0.52, 1.00) | (0.51, 1.00) |
| Effective spin $\chi_{\rm eff}$ | $0.02^{+0.10}_{-0.03}$ | $0.01^{+0.07}_{-0.02}$ | $0.01^{+0.07}_{-0.02}$ |
| Primary dimensionless spin $\chi_1$ | (0.00, 0.25) | (0.00, 0.25) | (0.00, 0.25) |
| Secondary dimensionless spin $\chi_2$ | (0.00, 0.39) | (0.00, 0.36) | (0.00, 0.35) |
| Tidal deformability $\tilde{\Lambda}$ with flat prior | (0, 730) | (0, 630) | (0, 640) |
| **Low-spin prior, $\chi_i \leq 0.05$** | **TaylorF2** | **SEOBNRT** | **PhenomDNRT** |
| Binary inclination $\theta_{JN}$ | $146^{+24}_{-28}$ deg | $146^{+24}_{-28}$ deg | $147^{+24}_{-28}$ deg |
| Binary inclination $\theta_{JN}$ using EM distance constraint [108] | $149^{+13}_{-10}$ deg | $152^{+14}_{-11}$ deg | $151^{+14}_{-10}$ deg |
| Detector-frame chirp mass $\mathcal{M}^{\rm det}$ | $1.1975^{+0.0001}_{-0.0001}$ $M_\odot$ | $1.1976^{+0.0001}_{-0.0001}$ $M_\odot$ | $1.1975^{+0.0001}_{-0.0001}$ $M_\odot$ |
| Chirp mass $\mathcal{M}$ | $1.186^{+0.001}_{-0.001}$ $M_\odot$ | $1.186^{+0.001}_{-0.001}$ $M_\odot$ | $1.186^{+0.001}_{-0.001}$ |
| Primary mass $m_1$ | $(1.36, 1.61)$ $M_\odot$ | $(1.36, 1.59)$ $M_\odot$ | $(1.36, 1.60)$ $M_\odot$ |
| Secondary mass $m_2$ | $(1.16, 1.36)$ $M_\odot$ | $(1.17, 1.36)$ $M_\odot$ | $(1.17, 1.36)$ $M_\odot$ |
| Total mass $m$ | $2.73^{+0.05}_{-0.01}$ $M_\odot$ | $2.73^{+0.04}_{-0.01}$ $M_\odot$ | $2.73^{+0.04}_{-0.01}$ $M_\odot$ |
| Mass ratio $q$ | (0.72, 1.00) | (0.74, 1.00) | (0.73, 1.00) |
| Effective spin $\chi_{\rm eff}$ | $0.00^{+0.02}_{-0.01}$ | $0.00^{+0.02}_{-0.01}$ | $0.00^{+0.02}_{-0.01}$ |
| Primary dimensionless spin $\chi_1$ | (0.00, 0.02) | (0.00, 0.02) | (0.00, 0.02) |
| Secondary dimensionless spin $\chi_2$ | (0.00, 0.02) | (0.00, 0.02) | (0.00, 0.02) |
| Tidal deformability $\tilde{\Lambda}$ with flat prior (symmetric/HPD) | $340^{+580}_{-240}/340^{+490}_{-290}$ | $280^{+490}_{-190}/280^{+410}_{-230}$ | $300^{+520}_{-190}/300^{+430}_{-230}$ |

TABLE V. Source properties for GW170817 produced using RapidPE for the additional waveform models SEOBNRv4T and TEOBResumS. Conventions are the same as in Table II.

| Low-spin prior, $\chi_i \leq 0.05$ | SEOBNRv4T | TEOBResumS | PhenomDNRT |
|---|---|---|---|
| Detector-frame chirp mass $\mathcal{M}^{\rm det}$ | $1.1975^{+0.0001}_{-0.0001}$ $M_\odot$ | $1.1975^{+0.0001}_{-0.0001}$ $M_\odot$ | $1.1975^{+0.0001}_{-0.0001}$ $M_\odot$ |
| Chirp mass $\mathcal{M}$ | $1.186^{+0.001}_{-0.001}$ $M_\odot$ | $1.186^{+0.001}_{-0.001}$ $M_\odot$ | $1.186^{+0.001}_{-0.001}$ |
| Primary mass $m_1$ | $(1.36, 1.56)$ $M_\odot$ | $(1.36, 1.53)$ $M_\odot$ | $(1.36, 1.57)$ $M_\odot$ |
| Secondary mass $m_2$ | $(1.19, 1.36)$ $M_\odot$ | $(1.22, 1.36)$ $M_\odot$ | $(1.19, 1.36)$ $M_\odot$ |
| Total mass $m$ | $2.73^{+0.04}_{-0.01}$ $M_\odot$ | $2.73^{+0.03}_{-0.01}$ $M_\odot$ | $2.73^{+0.04}_{-0.01}$ $M_\odot$ |
| Mass ratio $q$ | (0.76, 1.00) | (0.79, 1.00) | (0.76, 1.00) |
| Effective spin $\chi_{\rm eff}$ | $0.00^{+0.02}_{-0.01}$ | $0.00^{+0.01}_{-0.01}$ | $0.00^{+0.02}_{-0.01}$ |
| Primary dimensionless spin $\chi_1$ | (0.00, 0.03) | (0.00, 0.02) | (0.00, 0.03) |
| Secondary dimensionless spin $\chi_2$ | (0.00, 0.03) | (0.00, 0.03) | (0.00, 0.03) |
| Tidal deformability $\tilde{\Lambda}$ with flat prior (symmetric/HPD) | $280^{+430}_{-220}/280^{+280}_{-280}$ | $340^{+520}_{-260}/340^{+350}_{-330}$ | $310^{+510}_{-240}/310^{+380}_{-290}$ |

TABLE VI. Parameters used for the injected SEOBNRv4T waveform. The chosen masses and spins are consistent with the measured posteriors for GW170817. The tidal parameters are calculated from the mass and chosen EOS.

| Injection | $(m_1, m_2)$ $(M_\odot)$ | $(\chi_1, \chi_2)$ | EOS | $(\Lambda_1, \Lambda_2)$ | $\tilde{\Lambda}$ |
|---|---|---|---|---|---|
| i | (1.38, 1.37) | (0, 0) | APR4 | (275, 309) | 292 |
| ii | (1.68, 1.13) | (0, 0) | APR4 | (77, 973) | 303 |
| iii | (1.38, 1.37) | (0.04, 0) | APR4 | (275, 309) | 292 |
| iv | (1.38, 1.37) | (0, 0) | H | (1018, 1063) | 1040 |

## APPENDIX B: INJECTION AND RECOVERY STUDY

The reliability of the parameter estimation techniques used here was studied in detail for the first BBH detection by injecting state-of-the-art numerical waveform models into the data and verifying that the waveform templates correctly recover the injected parameters [115]. We perform a similar analysis for GW170817 by injecting a state-of-the-art BNS waveform model using parameters consistent with the data and then verifying that our waveform templates reliably recover the injected values. For BNS systems, we focus on the additional tidal parameters that are particularly sensitive to errors in the waveform models [138,168–171]. We use as our injected waveform model the time-domain aligned-spin SEOBNRv4T model [81,139] discussed in Sec. III D. The version used in this study did not include the spin-induced quadrupole moment, but later implementations of SEOBNRv4T such as the one used for the results in Table V include this effect.

We inject SEOBNRv4T with the following parameters in Table VI, which are consistent with the measured posterior for GW170817: (i) an approximately equal-mass, nonspinning case, (ii) an unequal-mass-ratio ($q = 0.67$), nonspinning case, and (iii) an approximately equal-mass case with a small spin for the primary star. For these systems, we choose a reference EOS that is near the peak of the tidal parameter $\tilde{\Lambda}$ in Fig. 11, APR4, from which we calculate the tidal parameters $\Lambda_1$ and $\Lambda_2$. Finally, we also choose (iv) a stiffer parametrized EOS that is sometimes used in NR simulations, H [172], which is near the maximum allowed value of $\tilde{\Lambda}$. We use the high-spin prior ($\chi_i \leq 0.89$) and the three aligned-spin waveform models (TaylorF2, PhenomDNRT, and SEOBNRT) as templates. In all four cases, we use the same PSD used in the analysis of GW170817 and inject the waveform with a network SNR of 32, consistent with GW170817. While the PSD is nonzero, we inject these waveforms into a zero-noise realization of the data. In other words, we assume the noise is zero at all frequencies. This has the advantage of making the results independent on possible large fluctuations of the Gaussian noise. Results obtained with zero noise are statistically equivalent to averaging results obtained with Gaussian noise over a large number of random realizations of Gaussian noise, and they are routinely presented in gravitational-wave literature [115,173–176].

We show in Fig. 14 the recovered tidal parameter $\tilde{\Lambda}$ when using the soft APR4 EOS. The posteriors for the three templates are peaked near the injected value of $\tilde{\Lambda}$. As with Fig. 11, the 90% upper limits are nearly the same for the two waveforms that use the NRTidal description, while the 90% upper limit for TaylorF2 is about 100 units higher.

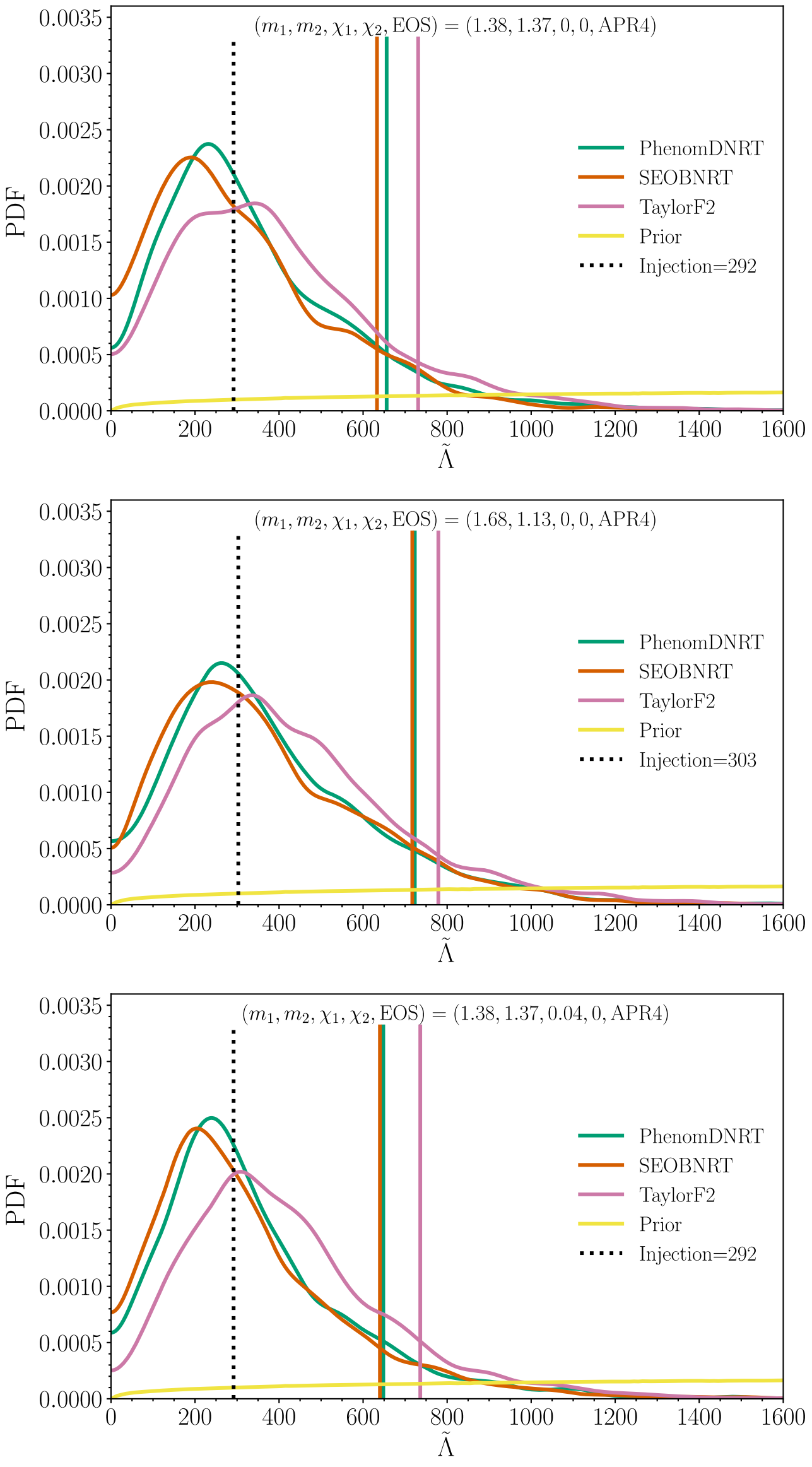

FIG. 14. Marginalized PDF of $\tilde{\Lambda}$ for the three aligned-spin waveform models using the high-spin prior of $\chi_i < 0.89$. As in Fig. 11, the PDF is reweighted by the prior. The SEOBNRv4T model was injected into zero-noise data with a network SNR of 32. The injected tidal parameter shown by the dotted vertical line was calculated with the APR4 EOS. Top panel: Approximately equal mass and nonspinning. Middle panel: Unequal mass and nonspinning. Bottom panel: Approximately equal mass and primary component spinning. Solid vertical lines represent the 90% upper limit for each waveform.

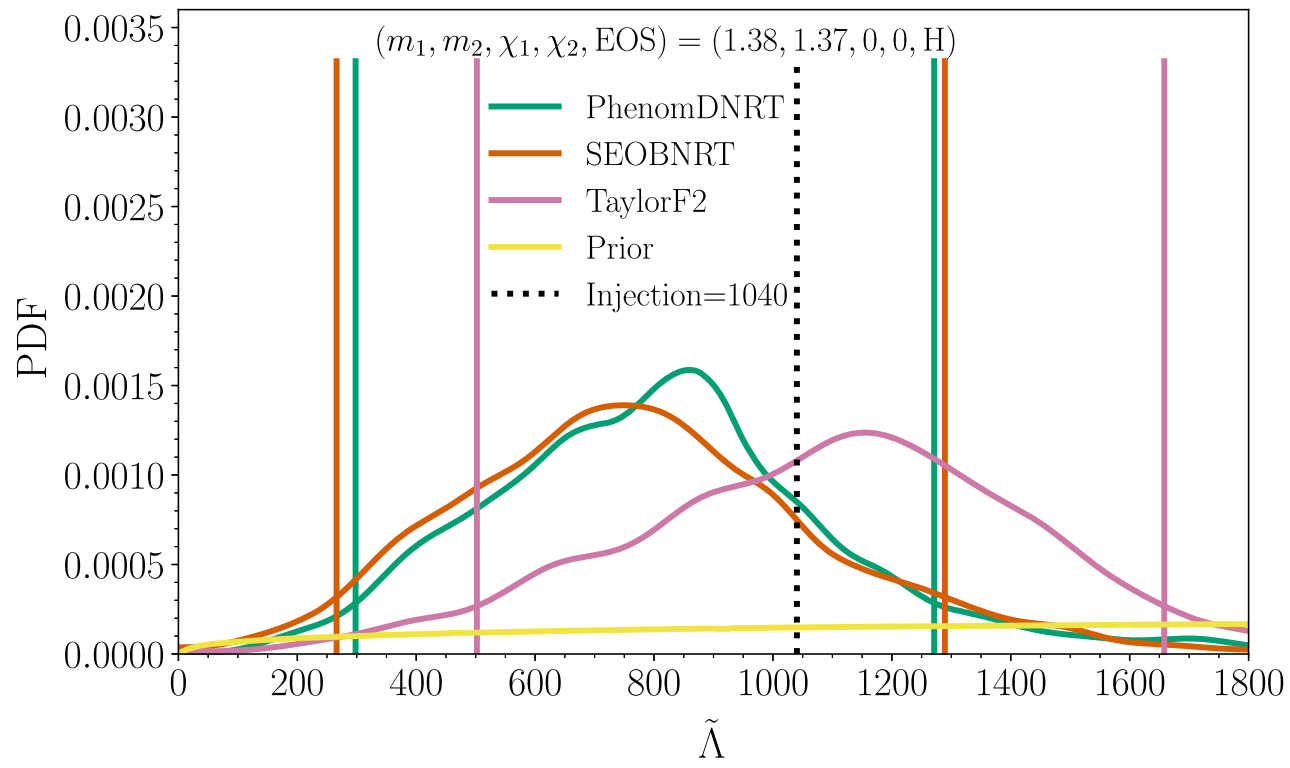

FIG. 15. Same as Fig. 14 but with injection (iv) using the H EOS. Solid vertical lines represent the 90% HPD credible interval for each waveform.

Because the TaylorF2 tidal effect is smaller than that for the NRTidal models, the TaylorF2 model will estimate a larger tidal parameter to compensate for the smaller tidal effect, cf. Fig. 2.

In Fig. 15, we show the recovered tidal parameter using the stiffer H EOS. The width of the posteriors, distance of the peaks from the injected value, and spread in the 90% credible intervals between the waveform models are larger than in Fig. 14, indicating that the statistical error and systematic waveform errors scale with the true tidal parameter. As with the APR4 injections, the credible interval for the NRTidal waveforms agree fairly well with each other, while the credible interval for the TaylorF2 waveform is about 400 units larger.

For GW170817 with a network SNR of 32, waveform systematic errors are important but do not dominate over statistical errors. However, as the detectors improve and results from multiple BNS observations are combined, the statistical errors will decrease. In this case, systematic waveform errors may become the dominant source of error, and improved waveform modeling will be needed.

B. P. Abbott,[1] R. Abbott,[1] T. D. Abbott,[2] F. Acernese,[3,4] K. Ackley,[5] C. Adams,[6] T. Adams,[7] P. Addesso,[8] R. X. Adhikari,[1] V. B. Adya,[9,10] C. Affeldt,[9,10] B. Agarwal,[11] M. Agathos,[12] K. Agatsuma,[13] N. Aggarwal,[14] O. D. Aguiar,[15] L. Aiello,[16,17] A. Ain,[18] P. Ajith,[19] B. Allen,[9,20,10] G. Allen,[11] A. Allocca,[21,22] M. A. Aloy,[23] P. A. Altin,[24] A. Amato,[25] A. Ananyeva,[1] S. B. Anderson,[1] W. G. Anderson,[20] S. V. Angelova,[26] S. Antier,[27] S. Appert,[1] K. Arai,[1] M. C. Araya,[1] J. S. Areeda,[28] M. Arène,[29] N. Arnaud,[30,31] K. G. Arun,[32] S. Ascenzi,[33,34] G. Ashton,[5] M. Ast,[35] S. M. Aston,[6] P. Astone,[36] D. V. Atallah,[37] F. Aubin,[38] P. Aufmuth,[10] C. Aulbert,[9] K. AultONeal,[39] C. Austin,[2] A. Avila-Alvarez,[28] S. Babak,[40,29] P. Bacon,[29] F. Badaracco,[16,17] M. K. M. Bader,[13] S. Bae,[41] P. T. Baker,[42] F. Baldaccini,[43,44] G. Ballardin,[31] S. W. Ballmer,[45] S. Banagiri,[46]

J. C. Barayoga,[1] S. E. Barclay,[47] B. C. Barish,[1] D. Barker,[48] K. Barkett,[49] S. Barnum,[14] F. Barone,[3,4] B. Barr,[47] L. Barsotti,[14] M. Barsuglia,[29] D. Barta,[50] J. Bartlett,[48] I. Bartos,[51] R. Bassiri,[52] A. Basti,[21,22] J. C. Batch,[48] M. Bawaj,[53,44] J. C. Bayley,[47] M. Bazzan,[54,55] B. Bécsy,[56] C. Beer,[9] M. Bejger,[57] I. Belahcene,[30] A. S. Bell,[47] D. Beniwal,[58] M. Bensch,[9,10] B. K. Berger,[1] G. Bergmann,[9,10] S. Bernuzzi,[59,60] J. J. Bero,[61] C. P. L. Berry,[62] D. Bersanetti,[63] A. Bertolini,[13] J. Betzwieser,[6] R. Bhandare,[64] I. A. Bilenko,[65] S. A. Bilgili,[42] G. Billingsley,[1] C. R. Billman,[51] J. Birch,[6] R. Birney,[26] O. Birnholtz,[61] S. Biscans,[1,14] S. Biscoveanu,[5] A. Bisht,[9,10] M. Bitossi,[31,22] M. A. Bizouard,[30] J. K. Blackburn,[1] J. Blackman,[49] C. D. Blair,[6] D. G. Blair,[66] R. M. Blair,[48] S. Bloemen,[67] O. Bock,[9] N. Bode,[9,10] M. Boer,[68] Y. Boetzel,[69] G. Bogaert,[68] A. Bohe,[40] F. Bondu,[70] E. Bonilla,[52] R. Bonnand,[38] P. Booker,[9,10] B. A. Boom,[13] C. D. Booth,[37] R. Bork,[1] V. Boschi,[31] S. Bose,[71,18] K. Bossie,[6] V. Bossilkov,[66] J. Bosveld,[66] Y. Bouffanais,[29] A. Bozzi,[31] C. Bradaschia,[22] P. R. Brady,[20] A. Bramley,[6] M. Branchesi,[16,17] J. E. Brau,[72] T. Briant,[73] F. Brighenti,[74,75] A. Brillet,[68] M. Brinkmann,[9,10] V. Brisson,[30,a] P. Brockill,[20] A. F. Brooks,[1] D. D. Brown,[58] S. Brunett,[1] C. C. Buchanan,[2] A. Buikema,[14] T. Bulik,[76] H. J. Bulten,[77,13] A. Buonanno,[40,78] D. Buskulic,[38] C. Buy,[29] R. L. Byer,[52] M. Cabero,[9] L. Cadonati,[79] G. Cagnoli,[25,80] C. Cahillane,[1] J. Calderón Bustillo,[79] T. A. Callister,[1] E. Calloni,[81,4] J. B. Camp,[82] M. Canepa,[83,63] P. Canizares,[67] K. C. Cannon,[84] H. Cao,[58] J. Cao,[85] C. D. Capano,[9] E. Capocasa,[29] F. Carbognani,[31] S. Caride,[86] M. F. Carney,[87] G. Carullo,[21] J. Casanueva Diaz,[22] C. Casentini,[33,34] S. Caudill,[13,20] M. Cavaglià,[88] F. Cavalier,[30] R. Cavalieri,[31] G. Cella,[22] C. B. Cepeda,[1] P. Cerdá-Durán,[23] G. Cerretani,[21,22] E. Cesarini,[89,34] O. Chaibi,[68] S. J. Chamberlin,[90] M. Chan,[47] S. Chao,[91] P. Charlton,[92] E. Chase,[93] E. Chassande-Mottin,[29] D. Chatterjee,[20] K. Chatziioannou,[94] B. D. Cheeseboro,[42] H. Y. Chen,[95] X. Chen,[66] Y. Chen,[49] H.-P. Cheng,[51] H. Y. Chia,[51] A. Chincarini,[63] A. Chiummo,[31] T. Chmiel,[87] H. S. Cho,[96] M. Cho,[78] J. H. Chow,[24] N. Christensen,[97,68] Q. Chu,[66] A. J. K. Chua,[49] S. Chua,[73] K. W. Chung,[98] S. Chung,[66] G. Ciani,[54,55,51] A. A. Ciobanu,[58] R. Ciolfi,[99,100] F. Cipriano,[68] C. E. Cirelli,[52] A. Cirone,[83,63] F. Clara,[48] J. A. Clark,[79] P. Clearwater,[101] F. Cleva,[68] C. Cocchieri,[88] E. Coccia,[16,17] P.-F. Cohadon,[73] D. Cohen,[30] A. Colla,[102,36] C. G. Collette,[103] C. Collins,[62] L. R. Cominsky,[104] M. Constancio Jr.,[15] L. Conti,[55] S. J. Cooper,[62] P. Corban,[6] T. R. Corbitt,[2] I. Cordero-Carrión,[105] K. R. Corley,[106] N. Cornish,[107] A. Corsi,[86] S. Cortese,[31] C. A. Costa,[15] R. Cotesta,[40] M. W. Coughlin,[1] S. B. Coughlin,[37,93] J.-P. Coulon,[68] S. T. Countryman,[106] P. Couvares,[1] P. B. Covas,[108] E. E. Cowan,[79] D. M. Coward,[66] M. J. Cowart,[6] D. C. Coyne,[1] R. Coyne,[109] J. D. E. Creighton,[20] T. D. Creighton,[110] J. Cripe,[2] S. G. Crowder,[111] T. J. Cullen,[2] A. Cumming,[47] L. Cunningham,[47] E. Cuoco,[31] T. Dal Canton,[82] G. Dálya,[56] S. L. Danilishin,[10,9] S. D'Antonio,[34] K. Danzmann,[9,10] A. Dasgupta,[112] C. F. Da Silva Costa,[51] V. Dattilo,[31] I. Dave,[64] M. Davier,[30] D. Davis,[45] E. J. Daw,[113] B. Day,[79] D. DeBra,[52] M. Deenadayalan,[18] J. Degallaix,[25] M. De Laurentis,[81,4] S. Deléglise,[73] W. Del Pozzo,[21,22] N. Demos,[14] T. Denker,[9,10] T. Dent,[9] R. De Pietri,[59,60] J. Derby,[28] V. Dergachev,[9] R. De Rosa,[81,4] C. De Rossi,[25,31] R. DeSalvo,[114] O. de Varona,[9,10] S. Dhurandhar,[18] M. C. Díaz,[110] T. Dietrich,[13,40] L. Di Fiore,[4] M. Di Giovanni,[115,100] T. Di Girolamo,[81,4] A. Di Lieto,[21,22] B. Ding,[103] S. Di Pace,[102,36] I. Di Palma,[116,36] F. Di Renzo,[21,22] A. Dmitriev,[62] Z. Doctor,[95] V. Dolique,[25] F. Donovan,[14] K. L. Dooley,[37,88] S. Doravari,[9,10] I. Dorrington,[37] M. Dovale Álvarez,[62] T. P. Downes,[20] M. Drago,[9,16,17] C. Dreissigacker,[9,10] J. C. Driggers,[48] Z. Du,[85] R. Dudi,[37] P. Dupej,[47] S. E. Dwyer,[48] P. J. Easter,[5] T. B. Edo,[113] M. C. Edwards,[97] A. Effler,[6] H.-B. Eggenstein,[9,10] P. Ehrens,[1] J. Eichholz,[1] S. S. Eikenberry,[51] M. Eisenmann,[38] R. A. Eisenstein,[14] R. C. Essick,[95] H. Estelles,[108] D. Estevez,[38] Z. B. Etienne,[42] T. Etzel,[1] M. Evans,[14] T. M. Evans,[6] V. Fafone,[33,34,16] H. Fair,[45] S. Fairhurst,[37] X. Fan,[85] S. Farinon,[63] B. Farr,[72] W. M. Farr,[62] E. J. Fauchon-Jones,[37] M. Favata,[117] M. Fays,[37] C. Fee,[87] H. Fehrmann,[9] J. Feicht,[1] M. M. Fejer,[52] F. Feng,[29] A. Fernandez-Galiana,[14] I. Ferrante,[21,22] E. C. Ferreira,[15] F. Ferrini,[31] F. Fidecaro,[21,22] I. Fiori,[31] D. Fiorucci,[29] M. Fishbach,[95] R. P. Fisher,[45] J. M. Fishner,[14] M. Fitz-Axen,[46] R. Flaminio,[38,118] M. Fletcher,[47] H. Fong,[94] J. A. Font,[23,119] P. W. F. Forsyth,[24] S. S. Forsyth,[79] J.-D. Fournier,[68] S. Frasca,[102,36] F. Frasconi,[22] Z. Frei,[56] A. Freise,[62] R. Frey,[72] V. Frey,[30] P. Fritschel,[14] V. V. Frolov,[6] P. Fulda,[51] M. Fyffe,[6] H. A. Gabbard,[47] B. U. Gadre,[18] S. M. Gaebel,[62] J. R. Gair,[120] L. Gammaitoni,[43] M. R. Ganija,[58] S. G. Gaonkar,[18] A. Garcia,[28] C. García-Quirós,[108] F. Garufi,[81,4] B. Gateley,[48] S. Gaudio,[39] G. Gaur,[121] V. Gayathri,[122] G. Gemme,[63] E. Genin,[31] A. Gennai,[22] D. George,[11] J. George,[64] L. Gergely,[123] V. Germain,[38] S. Ghonge,[79] Abhirup Ghosh,[19] Archisman Ghosh,[13] S. Ghosh,[20] B. Giacomazzo,[115,100] J. A. Giaime,[2,6] K. D. Giardina,[6] A. Giazotto,[22,b] K. Gill,[39] G. Giordano,[3,4] L. Glover,[114] E. Goetz,[48] R. Goetz,[51] B. Goncharov,[5] G. González,[2] J. M. Gonzalez Castro,[21,22] A. Gopakumar,[124] M. L. Gorodetsky,[65] S. E. Gossan,[1] M. Gosselin,[31] R. Gouaty,[38] A. Grado,[125,4] C. Graef,[47] M. Granata,[25] A. Grant,[47] S. Gras,[14] C. Gray,[48] G. Greco,[74,75] A. C. Green,[62] R. Green,[37] E. M. Gretarsson,[39] P. Groot,[67] H. Grote,[37] S. Grunewald,[40] P. Gruning,[30] G. M. Guidi,[74,75] H. K. Gulati,[112] X. Guo,[85] A. Gupta,[90] M. K. Gupta,[112] K. E. Gushwa,[1] E. K. Gustafson,[1] R. Gustafson,[126] O. Halim,[17,16] B. R. Hall,[71] E. D. Hall,[14]

E. Z. Hamilton,[37] H. F. Hamilton,[127] G. Hammond,[47] M. Haney,[69] M. M. Hanke,[9,10] J. Hanks,[48] C. Hanna,[90] M. D. Hannam,[37] O. A. Hannuksela,[98] J. Hanson,[6] T. Hardwick,[2] J. Harms,[16,17] G. M. Harry,[128] I. W. Harry,[40] M. J. Hart,[47] C.-J. Haster,[94] K. Haughian,[47] J. Healy,[61] A. Heidmann,[73] M. C. Heintze,[6] H. Heitmann,[68] P. Hello,[30] G. Hemming,[31] M. Hendry,[47] I. S. Heng,[47] J. Hennig,[47] A. W. Heptonstall,[1] F. J. Hernandez,[5] M. Heurs,[9,10] S. Hild,[47] T. Hinderer,[67] D. Hoak,[31] S. Hochheim,[9,10] D. Hofman,[25] N. A. Holland,[24] K. Holt,[6] D. E. Holz,[95] P. Hopkins,[37] C. Horst,[20] J. Hough,[47] E. A. Houston,[47] E. J. Howell,[66] A. Hreibi,[68] E. A. Huerta,[11] D. Huet,[30] B. Hughey,[39] M. Hulko,[1] S. Husa,[108] S. H. Huttner,[47] T. Huynh-Dinh,[6] A. Iess,[33,34] N. Indik,[9] C. Ingram,[58] R. Inta,[86] G. Intini,[102,36] H. N. Isa,[47] J.-M. Isac,[73] M. Isi,[1] B. R. Iyer,[19] K. Izumi,[129] T. Jacqmin,[73] K. Jani,[130] P. Jaranowski,[131] D. S. Johnson,[132] W. W. Johnson,[2] D. I. Jones,[133] R. Jones,[47] R. J. G. Jonker,[13] L. Ju,[66] J. Junker,[9,10] C. V. Kalaghatgi,[37] V. Kalogera,[134] B. Kamai,[135] S. Kandhasamy,[136] G. Kang,[41] J. B. Kanner,[135] S. J. Kapadia,[137] S. Karki,[138] K. S. Karvinen,[9,10] M. Kasprzack,[2] W. Kastaun,[9] M. Katolik,[132] S. Katsanevas,[31] E. Katsavounidis,[139] W. Katzman,[136] S. Kaufer,[9,10] K. Kawabe,[129] N. V. Keerthana,[18] F. Kéfélian,[68] D. Keitel,[47] A. J. Kemball,[132] R. Kennedy,[113] J. S. Key,[140] F. Y. Khalili,[65] B. Khamesra,[130] H. Khan,[141] I. Khan,[16,34] S. Khan,[9] Z. Khan,[112] E. A. Khazanov,[142] N. Kijbunchoo,[24] Chunglee Kim,[143] J. C. Kim,[144] K. Kim,[98] W. Kim,[58] W. S. Kim,[145] Y.-M. Kim,[146] E. J. King,[58] P. J. King,[129] M. Kinley-Hanlon,[128] R. Kirchhoff,[9,10] J. S. Kissel,[129] L. Kleybolte,[35] S. Klimenko,[147] T. D. Knowles,[148] P. Koch,[9,10] S. M. Koehlenbeck,[9,10] S. Koley,[13] V. Kondrashov,[135] A. Kontos,[139] M. Korobko,[35] W. Z. Korth,[135] I. Kowalska,[76] D. B. Kozak,[135] C. Krämer,[9] V. Kringel,[9,10] B. Krishnan,[9] A. Królak,[149,150] G. Kuehn,[9,10] P. Kumar,[151] R. Kumar,[112] S. Kumar,[19] L. Kuo,[91] A. Kutynia,[149] S. Kwang,[137] B. D. Lackey,[40] K. H. Lai,[98] M. Landry,[129] P. Landry,[152] R. N. Lang,[153] J. Lange,[154] B. Lantz,[155] R. K. Lanza,[139] A. Lartaux-Vollard,[30] P. D. Lasky,[5] M. Laxen,[136] A. Lazzarini,[135] C. Lazzaro,[55] P. Leaci,[156,36] S. Leavey,[9,10] C. H. Lee,[96] H. K. Lee,[157] H. M. Lee,[143] H. W. Lee,[144] K. Lee,[47] J. Lehmann,[9,10] A. Lenon,[148] M. Leonardi,[9,10,118] N. Leroy,[30] N. Letendre,[38] Y. Levin,[5] J. Li,[85] T. G. F. Li,[98] X. Li,[158] S. D. Linker,[159] T. B. Littenberg,[160] J. Liu,[66] X. Liu,[137] R. K. L. Lo,[98] N. A. Lockerbie,[26] L. T. London,[37] A. Longo,[161,162] M. Lorenzini,[16,17] V. Loriette,[163] M. Lormand,[136] G. Losurdo,[22] J. D. Lough,[9,10] C. O. Lousto,[154] G. Lovelace,[141] H. Lück,[9,10] D. Lumaca,[33,34] A. P. Lundgren,[9] R. Lynch,[139] Y. Ma,[158] R. Macas,[37] S. Macfoy,[26] B. Machenschalk,[9] M. MacInnis,[139] D. M. Macleod,[37] I. Magaña Hernandez,[137] F. Magaña-Sandoval,[164] L. Magaña Zertuche,[165] R. M. Magee,[166] E. Majorana,[36] I. Maksimovic,[163] N. Man,[68] V. Mandic,[167] V. Mangano,[47] G. L. Mansell,[24] M. Manske,[137,24] M. Mantovani,[31] F. Marchesoni,[53,44] F. Marion,[38] S. Márka,[168] Z. Márka,[168] C. Markakis,[132] A. S. Markosyan,[155] A. Markowitz,[135] E. Maros,[135] A. Marquina,[105] F. Martelli,[169,75] L. Martellini,[68] I. W. Martin,[47] R. M. Martin,[170] D. V. Martynov,[139] K. Mason,[139] E. Massera,[113] A. Masserot,[38] T. J. Massinger,[135] M. Masso-Reid,[47] S. Mastrogiovanni,[156,36] A. Matas,[167] F. Matichard,[135,139] L. Matone,[168] N. Mavalvala,[139] N. Mazumder,[171] J. J. McCann,[66] R. McCarthy,[129] D. E. McClelland,[24] S. McCormick,[136] L. McCuller,[139] S. C. McGuire,[172] J. McIver,[135] D. J. McManus,[24] T. McRae,[24] S. T. McWilliams,[148] D. Meacher,[166] G. D. Meadors,[5] M. Mehmet,[9,10] J. Meidam,[13] E. Mejuto-Villa,[8] A. Melatos,[101] G. Mendell,[129] D. Mendoza-Gandara,[9,10] R. A. Mercer,[137] L. Mereni,[25] E. L. Merilh,[129] M. Merzougui,[68] S. Meshkov,[135] C. Messenger,[47] C. Messick,[166] R. Metzdorff,[73] P. M. Meyers,[167] H. Miao,[62] C. Michel,[25] H. Middleton,[101] E. E. Mikhailov,[173] L. Milano,[81,4] A. L. Miller,[147] A. Miller,[156,36] B. B. Miller,[134] J. Miller,[139] M. Millhouse,[174] J. Mills,[37] M. C. Milovich-Goff,[159] O. Minazzoli,[68,175] Y. Minenkov,[34] J. Ming,[9,10] C. Mishra,[176] S. Mitra,[18] V. P. Mitrofanov,[65] G. Mitselmakher,[147] R. Mittleman,[139] D. Moffa,[177] K. Mogushi,[165] M. Mohan,[31] S. R. P. Mohapatra,[139] M. Montani,[169,75] C. J. Moore,[12] D. Moraru,[129] G. Moreno,[129] S. Morisaki,[178] B. Mours,[38] C. M. Mow-Lowry,[62] G. Mueller,[147] A. W. Muir,[37] Arunava Mukherjee,[9,10] D. Mukherjee,[137] S. Mukherjee,[179] N. Mukund,[18] A. Mullavey,[136] J. Munch,[58] E. A. Muñiz,[164] M. Muratore,[180] P. G. Murray,[47] A. Nagar,[89,181,182] K. Napier,[130] I. Nardecchia,[33,34] L. Naticchioni,[156,36] R. K. Nayak,[183] J. Neilson,[159] G. Nelemans,[67,13] T. J. N. Nelson,[136] M. Nery,[9,10] A. Neunzert,[184] L. Nevin,[135] J. M. Newport,[128] K. Y. Ng,[139] S. Ng,[58] P. Nguyen,[138] T. T. Nguyen,[24] D. Nichols,[67] A. B. Nielsen,[9] S. Nissanke,[67,13] A. Nitz,[9] F. Nocera,[31] D. Nolting,[136] C. North,[37] L. K. Nuttall,[37] M. Obergaulinger,[23] J. Oberling,[129] B. D. O'Brien,[147] G. D. O'Dea,[159] G. H. Ogin,[185] J. J. Oh,[145] S. H. Oh,[145] F. Ohme,[9] H. Ohta,[178] M. A. Okada,[15] M. Oliver,[108] P. Oppermann,[9,10] Richard J. Oram,[136] B. O'Reilly,[136] R. Ormiston,[167] L. F. Ortega,[147] R. O'Shaughnessy,[154] S. Ossokine,[40] D. J. Ottaway,[58] H. Overmier,[136] B. J. Owen,[186] A. E. Pace,[166] G. Pagano,[21,22] J. Page,[160] M. A. Page,[66] A. Pai,[122] S. A. Pai,[64] J. R. Palamos,[138] O. Palashov,[142] C. Palomba,[36] A. Pal-Singh,[35] Howard Pan,[91] Huang-Wei Pan,[91] B. Pang,[158] P. T. H. Pang,[98] C. Pankow,[134] F. Pannarale,[37] B. C. Pant,[64] F. Paoletti,[22] A. Paoli,[31] M. A. Papa,[9,137,10] A. Parida,[18] W. Parker,[136] D. Pascucci,[47] A. Pasqualetti,[31] R. Passaquieti,[21,22] D. Passuello,[22] M. Patil,[150] B. Patricelli,[187,22] B. L. Pearlstone,[47] C. Pedersen,[37] M. Pedraza,[135] R. Pedurand,[25,188] L. Pekowsky,[164]

A. Pele,[136] S. Penn,[189] C. J. Perez,[129] A. Perreca,[115,100] L. M. Perri,[134] H. P. Pfeiffer,[94,40] M. Phelps,[47] K. S. Phukon,[18] O. J. Piccinni,[156,36] M. Pichot,[68] F. Piergiovanni,[169,75] V. Pierro,[8] G. Pillant,[31] L. Pinard,[25] I. M. Pinto,[8] M. Pirello,[129] M. Pitkin,[47] R. Poggiani,[21,22] P. Popolizio,[31] E. K. Porter,[29] L. Possenti,[190,75] A. Post,[9] J. Powell,[191] J. Prasad,[18] J. W. W. Pratt,[180] G. Pratten,[108] V. Predoi,[37] T. Prestegard,[137] M. Principe,[8] S. Privitera,[40] G. A. Prodi,[115,100] L. G. Prokhorov,[65] O. Puncken,[9,10] M. Punturo,[44] P. Puppo,[36] M. Pürrer,[40] H. Qi,[137] V. Quetschke,[179] E. A. Quintero,[135] R. Quitzow-James,[138] F. J. Raab,[129] D. S. Rabeling,[24] H. Radkins,[129] P. Raffai,[56] S. Raja,[64] C. Rajan,[64] B. Rajbhandari,[186] M. Rakhmanov,[179] K. E. Ramirez,[179] A. Ramos-Buades,[108] Javed Rana,[18] P. Rapagnani,[156,36] V. Raymond,[37] M. Razzano,[21,22] J. Read,[141] T. Regimbau,[68,38] L. Rei,[63] S. Reid,[26] D. H. Reitze,[135,147] W. Ren,[132] F. Ricci,[156,36] P. M. Ricker,[132] G. Riemenschneider,[181,192] K. Riles,[184] M. Rizzo,[154] N. A. Robertson,[135,47] R. Robie,[47] F. Robinet,[30] T. Robson,[174] A. Rocchi,[34] L. Rolland,[38] J. G. Rollins,[135] V. J. Roma,[138] R. Romano,[3,4] C. L. Romel,[129] J. H. Romie,[136] D. Rosińska,[193,57] M. P. Ross,[194] S. Rowan,[47] A. Rüdiger,[9,10] P. Ruggi,[31] G. Rutins,[195] K. Ryan,[129] S. Sachdev,[135] T. Sadecki,[129] M. Sakellariadou,[196] L. Salconi,[31] M. Saleem,[122] F. Salemi,[9] A. Samajdar,[183,13] L. Sammut,[5] L. M. Sampson,[134] E. J. Sanchez,[135] L. E. Sanchez,[135] N. Sanchis-Gual,[23] V. Sandberg,[129] J. R. Sanders,[164] N. Sarin,[5] B. Sassolas,[25] B. S. Sathyaprakash,[166,37] P. R. Saulson,[164] O. Sauter,[184] R. L. Savage,[129] A. Sawadsky,[35] P. Schale,[138] M. Scheel,[158] J. Scheuer,[134] P. Schmidt,[67] R. Schnabel,[35] R. M. S. Schofield,[138] A. Schönbeck,[35] E. Schreiber,[9,10] D. Schuette,[9,10] B. W. Schulte,[9,10] B. F. Schutz,[37,9] S. G. Schwalbe,[180] J. Scott,[47] S. M. Scott,[24] E. Seidel,[132] D. Sellers,[136] A. S. Sengupta,[197] D. Sentenac,[31] V. Sequino,[33,34,16] A. Sergeev,[142] Y. Setyawati,[9] D. A. Shaddock,[24] T. J. Shaffer,[129] A. A. Shah,[160] M. S. Shahriar,[134] M. B. Shaner,[159] L. Shao,[40] B. Shapiro,[155] P. Shawhan,[198] H. Shen,[132] D. H. Shoemaker,[139] D. M. Shoemaker,[130] K. Siellez,[130] X. Siemens,[137] M. Sieniawska,[57] D. Sigg,[129] A. D. Silva,[15] L. P. Singer,[199] A. Singh,[9,10] A. Singhal,[16,36] A. M. Sintes,[108] B. J. J. Slagmolen,[24] T. J. Slaven-Blair,[66] B. Smith,[136] J. R. Smith,[141] R. J. E. Smith,[5] S. Somala,[200] E. J. Son,[145] B. Sorazu,[47] F. Sorrentino,[63] T. Souradeep,[18] A. P. Spencer,[47] A. K. Srivastava,[112] K. Staats,[180] M. Steinke,[9,10] J. Steinlechner,[35,47] S. Steinlechner,[35] D. Steinmeyer,[9,10] B. Steltner,[9,10] S. P. Stevenson,[191] D. Stocks,[155] R. Stone,[179] D. J. Stops,[62] K. A. Strain,[47] G. Stratta,[169,75] S. E. Strigin,[65] A. Strunk,[129] R. Sturani,[201] A. L. Stuver,[202] T. Z. Summerscales,[203] L. Sun,[101] S. Sunil,[112] J. Suresh,[18] P. J. Sutton,[37] B. L. Swinkels,[13] M. J. Szczepańczyk,[180] M. Tacca,[13] S. C. Tait,[47] C. Talbot,[5] D. Talukder,[138] D. B. Tanner,[147] M. Tápai,[123] A. Taracchini,[40] J. D. Tasson,[204] J. A. Taylor,[160] R. Taylor,[135] S. V. Tewari,[189] T. Theeg,[9,10] F. Thies,[9,10] E. G. Thomas,[62] M. Thomas,[136] P. Thomas,[129] K. A. Thorne,[136] E. Thrane,[5] S. Tiwari,[16,100] V. Tiwari,[37] K. V. Tokmakov,[26] K. Toland,[47] M. Tonelli,[21,22] Z. Tornasi,[47] A. Torres-Forné,[23] C. I. Torrie,[135] D. Töyrä,[62] F. Travasso,[31,44] G. Traylor,[136] J. Trinastic,[147] M. C. Tringali,[115,100] L. Trozzo,[205,22] K. W. Tsang,[13] M. Tse,[139] R. Tso,[158] D. Tsuna,[178] L. Tsukada,[178] D. Tuyenbayev,[179] K. Ueno,[137] D. Ugolini,[206] A. L. Urban,[135] S. A. Usman,[37] H. Vahlbruch,[9,10] G. Vajente,[135] G. Valdes,[2] N. van Bakel,[13] M. van Beuzekom,[13] J. F. J. van den Brand,[77,13] C. Van Den Broeck,[13,207] D. C. Vander-Hyde,[164] L. van der Schaaf,[13] J. V. van Heijningen,[13] A. A. van Veggel,[47] M. Vardaro,[54,55] V. Varma,[158] S. Vass,[135] M. Vasúth,[50] A. Vecchio,[62] G. Vedovato,[55] J. Veitch,[47] P. J. Veitch,[58] K. Venkateswara,[194] G. Venugopalan,[135] D. Verkindt,[38] F. Vetrano,[169,75] A. Viceré,[169,75] A. D. Viets,[137] S. Vinciguerra,[62] D. J. Vine,[195] J.-Y. Vinet,[68] S. Vitale,[139] T. Vo,[164] H. Vocca,[43,44] C. Vorvick,[129] S. P. Vyatchanin,[65] A. R. Wade,[135] L. E. Wade,[177] M. Wade,[177] R. Walet,[13] M. Walker,[141] L. Wallace,[135] S. Walsh,[137,9] G. Wang,[16,22] H. Wang,[62] J. Z. Wang,[184] W. H. Wang,[179] Y. F. Wang,[98] R. L. Ward,[24] J. Warner,[129] M. Was,[38] J. Watchi,[103] B. Weaver,[129] L.-W. Wei,[9,10] M. Weinert,[9,10] A. J. Weinstein,[135] R. Weiss,[139] F. Wellmann,[9,10] L. Wen,[66] E. K. Wessel,[132] P. Weßels,[9,10] J. Westerweck,[9] K. Wette,[24] J. T. Whelan,[154] B. F. Whiting,[147] C. Whittle,[139] D. Wilken,[9,10] D. Williams,[47] R. D. Williams,[135] A. R. Williamson,[154,67] J. L. Willis,[135,208] B. Willke,[9,10] M. H. Wimmer,[9,10] W. Winkler,[9,10] C. C. Wipf,[135] H. Wittel,[9,10] G. Woan,[47] J. Woehler,[9,10] J. K. Wofford,[154] W. K. Wong,[98] J. Worden,[129] J. L. Wright,[47] D. S. Wu,[9,10] D. M. Wysocki,[154] S. Xiao,[135] W. Yam,[139] H. Yamamoto,[135] C. C. Yancey,[198] L. Yang,[209] M. J. Yap,[24] M. Yazback,[147] Hang Yu,[139] Haocun Yu,[139] M. Yvert,[38] A. Zadrożny,[149] M. Zanolin,[180] T. Zelenova,[31] J.-P. Zendri,[55] M. Zevin,[134] J. Zhang,[66] L. Zhang,[135] M. Zhang,[173] T. Zhang,[47] Y.-H. Zhang,[9,10] C. Zhao,[66] M. Zhou,[134] Z. Zhou,[134] S. J. Zhu,[9,10] X. J. Zhu,[5] A. B. Zimmerman,[94] Y. Zlochower,[154] M. E. Zucker,[135,139] and J. Zweizig[135]

(LIGO Scientific Collaboration and Virgo Collaboration)

[1]*LIGO, California Institute of Technology, Pasadena, California 91125, USA*
[2]*Louisiana State University, Baton Rouge, Louisiana 70803, USA*
[3]*Università di Salerno, Fisciano, I-84084 Salerno, Italy*

[4]INFN, Sezione di Napoli, Complesso Universitario di Monte S.Angelo, I-80126 Napoli, Italy
[5]OzGrav, School of Physics & Astronomy, Monash University, Clayton 3800, Victoria, Australia
[6]LIGO Livingston Observatory, Livingston, Louisiana 70754, USA
[7]Laboratoire d'Annecy de Physique des Particules (LAPP), Univ. Grenoble Alpes, Université Savoie Mont Blanc, CNRS/IN2P3, F-74941 Annecy, France
[8]University of Sannio at Benevento, I-82100 Benevento, Italy and INFN, Sezione di Napoli, I-80100 Napoli, Italy
[9]Max Planck Institute for Gravitational Physics (Albert Einstein Institute), D-30167 Hannover, Germany
[10]Leibniz Universität Hannover, D-30167 Hannover, Germany
[11]NCSA, University of Illinois at Urbana-Champaign, Urbana, Illinois 61801, USA
[12]University of Cambridge, Cambridge CB2 1TN, United Kingdom
[13]Nikhef, Science Park 105, 1098 XG Amsterdam, Netherlands
[14]LIGO, Massachusetts Institute of Technology, Cambridge, Massachusetts 02139, USA
[15]Instituto Nacional de Pesquisas Espaciais, 12227-010 São José dos Campos, São Paulo, Brazil
[16]Gran Sasso Science Institute (GSSI), I-67100 L'Aquila, Italy
[17]INFN, Laboratori Nazionali del Gran Sasso, I-67100 Assergi, Italy
[18]Inter-University Centre for Astronomy and Astrophysics, Pune 411007, India
[19]International Centre for Theoretical Sciences, Tata Institute of Fundamental Research, Bengaluru 560089, India
[20]University of Wisconsin-Milwaukee, Milwaukee, Wisconsin 53201, USA
[21]Università di Pisa, I-56127 Pisa, Italy
[22]INFN, Sezione di Pisa, I-56127 Pisa, Italy
[23]Departamento de Astronomía y Astrofísica, Universitat de València, E-46100 Burjassot, València, Spain
[24]OzGrav, Australian National University, Canberra, Australian Capital Territory 0200, Australia
[25]Laboratoire des Matériaux Avancés (LMA), CNRS/IN2P3, F-69622 Villeurbanne, France
[26]SUPA, University of Strathclyde, Glasgow G1 1XQ, United Kingdom
[27]LAL, Univ. Paris-Sud, CNRS/IN2P3, Université Paris-Saclay, F-91898 Orsay, France
[28]California State University Fullerton, Fullerton, California 92831, USA
[29]APC, AstroParticule et Cosmologie, Université Paris Diderot, CNRS/IN2P3, CEA/Irfu, Observatoire de Paris, Sorbonne Paris Cité, F-75205 Paris Cedex 13, France
[30]LAL, Univ. Paris-Sud, CNRS/IN2P3, Université Paris-Saclay, F-91898 Orsay, France
[31]European Gravitational Observatory (EGO), I-56021 Cascina, Pisa, Italy
[32]Chennai Mathematical Institute, Chennai 603103, India
[33]Università di Roma Tor Vergata, I-00133 Roma, Italy
[34]INFN, Sezione di Roma Tor Vergata, I-00133 Roma, Italy
[35]Universität Hamburg, D-22761 Hamburg, Germany
[36]INFN, Sezione di Roma, I-00185 Roma, Italy
[37]Cardiff University, Cardiff CF24 3AA, United Kingdom
[38]Laboratoire d'Annecy de Physique des Particules (LAPP), Univ. Grenoble Alpes, Université Savoie Mont Blanc, CNRS/IN2P3, F-74941 Annecy, France
[39]Embry-Riddle Aeronautical University, Prescott, Arizona 86301, USA
[40]Max Planck Institute for Gravitational Physics (Albert Einstein Institute), D-14476 Potsdam-Golm, Germany
[41]Korea Institute of Science and Technology Information, Daejeon 34141, Korea
[42]West Virginia University, Morgantown, West Virginia 26506, USA
[43]Università di Perugia, I-06123 Perugia, Italy
[44]INFN, Sezione di Perugia, I-06123 Perugia, Italy
[45]Syracuse University, Syracuse, New York 13244, USA
[46]University of Minnesota, Minneapolis, Minnesota 55455, USA
[47]SUPA, University of Glasgow, Glasgow G12 8QQ, United Kingdom
[48]LIGO Hanford Observatory, Richland, Washington 99352, USA
[49]Caltech CaRT, Pasadena, California 91125, USA
[50]Wigner RCP, RMKI, H-1121 Budapest, Konkoly Thege Miklós út 29-33, Hungary
[51]University of Florida, Gainesville, Florida 32611, USA
[52]Stanford University, Stanford, California 94305, USA
[53]Università di Camerino, Dipartimento di Fisica, I-62032 Camerino, Italy
[54]Università di Padova, Dipartimento di Fisica e Astronomia, I-35131 Padova, Italy
[55]INFN, Sezione di Padova, I-35131 Padova, Italy
[56]MTA-ELTE Astrophysics Research Group, Institute of Physics, Eötvös University, Budapest 1117, Hungary

[57]Nicolaus Copernicus Astronomical Center, Polish Academy of Sciences, 00-716, Warsaw, Poland
[58]OzGrav, University of Adelaide, Adelaide, South Australia 5005, Australia
[59]Dipartimento di Scienze Matematiche, Fisiche e Informatiche, Università di Parma, I-43124 Parma, Italy
[60]INFN, Sezione di Milano Bicocca, Gruppo Collegato di Parma, I-43124 Parma, Italy
[61]Rochester Institute of Technology, Rochester, New York 14623, USA
[62]University of Birmingham, Birmingham B15 2TT, United Kingdom
[63]INFN, Sezione di Genova, I-16146 Genova, Italy
[64]RRCAT, Indore, Madhya Pradesh 452013, India
[65]Faculty of Physics, Lomonosov Moscow State University, Moscow 119991, Russia
[66]OzGrav, University of Western Australia, Crawley, Western Australia 6009, Australia
[67]Department of Astrophysics/IMAPP, Radboud University Nijmegen, P.O. Box 9010, 6500 GL Nijmegen, Netherlands
[68]Artemis, Université Côte d'Azur, Observatoire Côte d'Azur, CNRS, CS 34229, F-06304 Nice Cedex 4, France
[69]Physik-Institut, University of Zurich, Winterthurerstrasse 190, 8057 Zurich, Switzerland
[70]Univ Rennes, CNRS, Institut FOTON—UMR6082, F-3500 Rennes, France
[71]Washington State University, Pullman, Washington 99164, USA
[72]University of Oregon, Eugene, Oregon 97403, USA
[73]Laboratoire Kastler Brossel, Sorbonne Université, CNRS, ENS-Université PSL, Collège de France, F-75005 Paris, France
[74]Università degli Studi di Urbino "Carlo Bo," I-61029 Urbino, Italy
[75]INFN, Sezione di Firenze, I-50019 Sesto Fiorentino, Firenze, Italy
[76]Astronomical Observatory Warsaw University, 00-478 Warsaw, Poland
[77]VU University Amsterdam, 1081 HV Amsterdam, Netherlands
[78]University of Maryland, College Park, Maryland 20742, USA
[79]School of Physics, Georgia Institute of Technology, Atlanta, Georgia 30332, USA
[80]Université Claude Bernard Lyon 1, F-69622 Villeurbanne, France
[81]Università di Napoli 'Federico II,' Complesso Universitario di Monte S.Angelo, I-80126 Napoli, Italy
[82]NASA Goddard Space Flight Center, Greenbelt, Maryland 20771, USA
[83]Dipartimento di Fisica, Università degli Studi di Genova, I-16146 Genova, Italy
[84]RESCEU, University of Tokyo, Tokyo, 113-0033, Japan
[85]Tsinghua University, Beijing 100084, China
[86]Texas Tech University, Lubbock, Texas 79409, USA
[87]Kenyon College, Gambier, Ohio 43022, USA
[88]The University of Mississippi, University, Mississippi 38677, USA
[89]Museo Storico della Fisica e Centro Studi e Ricerche "Enrico Fermi", I-00184 Roma, Italyrico Fermi, I-00184 Roma, Italy
[90]The Pennsylvania State University, University Park, Pennsylvania 16802, USA
[91]National Tsing Hua University, Hsinchu City, 30013 Taiwan, Republic of China
[92]Charles Sturt University, Wagga Wagga, New South Wales 2678, Australia
[93]Center for Interdisciplinary Exploration & Research in Astrophysics (CIERA), Northwestern University, Evanston, Illinois 60208, USA
[94]Canadian Institute for Theoretical Astrophysics, University of Toronto, Toronto, Ontario M5S 3H8, Canada
[95]University of Chicago, Chicago, Illinois 60637, USA
[96]Pusan National University, Busan 46241, Korea
[97]Carleton College, Northfield, Minnesota 55057, USA
[98]The Chinese University of Hong Kong, Shatin, NT, Hong Kong
[99]INAF, Osservatorio Astronomico di Padova, I-35122 Padova, Italy
[100]INFN, Trento Institute for Fundamental Physics and Applications, I-38123 Povo, Trento, Italy
[101]OzGrav, University of Melbourne, Parkville, Victoria 3010, Australia
[102]Università di Roma "La Sapienza," I-00185 Roma, Italy
[103]Université Libre de Bruxelles, Brussels 1050, Belgium
[104]Sonoma State University, Rohnert Park, California 94928, USA
[105]Departamento de Matemáticas, Universitat de València, E-46100 Burjassot, València, Spain
[106]Columbia University, New York, New York 10027, USA
[107]Montana State University, Bozeman, Montana 59717, USA
[108]Universitat de les Illes Balears, IAC3—IEEC, E-07122 Palma de Mallorca, Spain
[109]University of Rhode Island, 45 Upper College Rd, Kingston, Rhode Island 02881, USA

[110]The University of Texas Rio Grande Valley, Brownsville, Texas 78520, USA
[111]Bellevue College, Bellevue, Washington 98007, USA
[112]Institute for Plasma Research, Bhat, Gandhinagar 382428, India
[113]The University of Sheffield, Sheffield S10 2TN, United Kingdom
[114]California State University, Los Angeles, 5151 State University Drive, Los Angeles, California 90032, USA
[115]Università di Trento, Dipartimento di Fisica, I-38123 Povo, Trento, Italy
[116]Università di Roma "La Sapienza," I-00185 Roma, Italy
[117]Montclair State University, Montclair, New Jersey 07043, USA
[118]National Astronomical Observatory of Japan, 2-21-1 Osawa, Mitaka, Tokyo 181-8588, Japan
[119]Observatori Astronòmic, Universitat de València, E-46980 Paterna, València, Spain
[120]School of Mathematics, University of Edinburgh, Edinburgh EH9 3FD, United Kingdom
[121]University and Institute of Advanced Research, Koba Institutional Area, Gandhinagar Gujarat 382007, India
[122]Indian Institute of Technology, Main Gate Rd, IIT Area, Powai, Mumbai, Maharashtra 400076, India
[123]University of Szeged, Dóm tér 9, Szeged 6720, Hungary
[124]Tata Institute of Fundamental Research, Mumbai 400005, India
[125]INAF, Osservatorio Astronomico di Capodimonte, I-80131, Napoli, Italy
[126]University of Michigan, Ann Arbor, Michigan 48109, USA
[127]Abilene Christian University, Abilene, Texas 79699, USA
[128]American University, Washington, D.C. 20016, USA
[129]LIGO Hanford Observatory, Richland, Washington 99352, USA
[130]School of Physics, Georgia Institute of Technology, Atlanta, Georgia 30332, USA
[131]University of Białystok, 15-424 Białystok, Poland
[132]NCSA, University of Illinois at Urbana-Champaign, Urbana, Illinois 61801, USA
[133]University of Southampton, Southampton SO17 1BJ, United Kingdom
[134]Center for Interdisciplinary Exploration & Research in Astrophysics (CIERA), Northwestern University, Evanston, Illinois 60208, USA
[135]LIGO, California Institute of Technology, Pasadena, California 91125, USA
[136]LIGO Livingston Observatory, Livingston, Louisiana 70754, USA
[137]University of Wisconsin-Milwaukee, Milwaukee, Wisconsin 53201, USA
[138]University of Oregon, Eugene, Oregon 97403, USA
[139]LIGO, Massachusetts Institute of Technology, Cambridge, Massachusetts 02139, USA
[140]University of Washington Bothell, 18115 Campus Way NE, Bothell, Washington 98011, USA
[141]California State University Fullerton, Fullerton, California 92831, USA
[142]Institute of Applied Physics, Nizhny Novgorod, 603950, Russia
[143]Korea Astronomy and Space Science Institute, Daejeon 34055, Korea
[144]Inje University Gimhae, South Gyeongsang 50834, Korea
[145]National Institute for Mathematical Sciences, Daejeon 34047, Korea
[146]Ulsan National Institute of Science and Technology
[147]University of Florida, Gainesville, Florida 32611, USA
[148]West Virginia University, Morgantown, West Virginia 26506, USA
[149]NCBJ, 05-400 Świerk-Otwock, Poland
[150]Institute of Mathematics, Polish Academy of Sciences, 00656 Warsaw, Poland
[151]Cornell Universtiy
[152]University of Chicago, Chicago, Illinois 60637, USA
[153]Hillsdale College, Hillsdale, Michigan 49242, USA
[154]Rochester Institute of Technology, Rochester, New York 14623, USA
[155]Stanford University, Stanford, California 94305, USA
[156]Università di Roma 'La Sapienza,' I-00185 Roma, Italy
[157]Hanyang University, Seoul 04763, Korea
[158]Caltech CaRT, Pasadena, California 91125, USA
[159]California State University, Los Angeles, 5151 State University Dr, Los Angeles, California 90032, USA
[160]NASA Marshall Space Flight Center, Huntsville, Alabama 35811, USA
[161]Dipartimento di Fisica, Università degli Studi Roma Tre, I-00154 Roma, Italy
[162]INFN, Sezione di Roma Tre, I-00154 Roma, Italy
[163]ESPCI, CNRS, F-75005 Paris, France
[164]Syracuse University, Syracuse, New York 13244, USA
[165]The University of Mississippi, University, Mississipp 38677, USA
[166]The Pennsylvania State University, University Park, Pennsylvania 16802, USA

[167]*University of Minnesota, Minneapolis, Minnesota 55455, USA*
[168]*Columbia University, New York, New York 10027, USA*
[169]*Università degli Studi di Urbino 'Carlo Bo,' I-61029 Urbino, Italy*
[170]*Montclair State University, Montclair, New Jersey 07043, USA*
[171]*Washington State University, Pullman, Washington 99164, USA*
[172]*Southern University and A&M College, Baton Rouge, Louisiana 70813, USA*
[173]*College of William and Mary, Williamsburg, Virginia 23187, USA*
[174]*Montana State University, Bozeman, Montana 59717, USA*
[175]*Centre Scientifique de Monaco, 8 quai Antoine Ier, MC-98000, Monaco*
[176]*Indian Institute of Technology Madras, Chennai 600036, India*
[177]*Kenyon College, Gambier, Ohio 43022, USA*
[178]*RESCEU, University of Tokyo, Tokyo, 113-0033, Japan*
[179]*The University of Texas Rio Grande Valley, Brownsville, Texas 78520, USA*
[180]*Embry-Riddle Aeronautical University, Prescott, Arizona 86301, USA*
[181]*INFN Sezione di Torino, Via P. Giuria 1, I-10125 Torino, Italy*
[182]*Institut des Hautes Etudes Scientifiques, F-91440 Bures-sur-Yvette, France*
[183]*IISER-Kolkata, Mohanpur, West Bengal 741252, India*
[184]*University of Michigan, Ann Arbor, Michigan 48109, USA*
[185]*Whitman College, 345 Boyer Avenue, Walla Walla, Washington 99362 USA*
[186]*Texas Tech University, Lubbock, Texas 79409, USA*
[187]*Scuola Normale Superiore, Piazza dei Cavalieri 7, I-56126 Pisa, Italy*
[188]*Université de Lyon, F-69361 Lyon, France*
[189]*Hobart and William Smith Colleges, Geneva, New York 14456, USA*
[190]*Università degli Studi di Firenze, I-50121 Firenze, Italy*
[191]*OzGrav, Swinburne University of Technology, Hawthorn VIC 3122, Australia*
[192]*Dipartimento di Fisica, Università di Torino, via P. Giuria 1, I-10125 Torino, Italy*
[193]*Janusz Gil Institute of Astronomy, University of Zielona Góra, 65-265 Zielona Góra, Poland*
[194]*University of Washington, Seattle, Washington 98195, USA*
[195]*SUPA, University of the West of Scotland, Paisley PA1 2BE, United Kingdom*
[196]*King's College London, University of London, London WC2R 2LS, United Kingdom*
[197]*Indian Institute of Technology, Gandhinagar Ahmedabad Gujarat 382424, India*
[198]*University of Maryland, College Park, Maryland 20742, USA*
[199]*NASA Goddard Space Flight Center, Greenbelt, Maryland 20771, USA*
[200]*Indian Institute of Technology Hyderabad, Sangareddy, Khandi, Telangana 502285, India*
[201]*International Institute of Physics, Universidade Federal do Rio Grande do Norte, Natal RN 59078-970, Brazil*
[202]*Villanova University, 800 Lancaster Ave, Villanova, Pennsylvania 19085, USA*
[203]*Andrews University, Berrien Springs, Michigan 49104, USA*
[204]*Carleton College, Northfield, Minnesota55057, USA*
[205]*Università di Siena, I-53100 Siena, Italy*
[206]*Trinity University, San Antonio, Texas 78212, USA*
[207]*Van Swinderen Institute for Particle Physics and Gravity, University of Groningen, Nijenborgh 4, 9747 AG Groningen, Netherlands*
[208]*Abilene Christian University, Abilene, Texas 79699, USA*
[209]*Colorado State University, Fort Collins, Colorado 80523, USA*

[a]Deceased, February 2018.
[b]Deceased, November 2017.